\documentclass[10pt]{article}
\usepackage{hyperref}
\usepackage{float}
\usepackage{amssymb,amsmath,graphicx,latexsym,cite}

\begin{document}

\title{A Non-static Quantum inspired Spacetime in $f(R)$ Gravity:
Gravity's Rainbow}

\author{Prabir Rudra \thanks{prudra.math@gmail.com, rudra@associates.iucaa.in}\\\emph{\small{Department of Mathematics, Asutosh College,
Kolkata-700 026, India.}}}

\maketitle

\textbf{The author dedicates this work to the memory of the people
who died due to COVID-19}

\begin{abstract}
In this note we explore a non-static spacetime in quantum regime
in the background of $f(R)$ gravity. The time dependent Vaidya
metric which represents the spacetime of a radiating body like
star is studied in an energy dependent gravity's rainbow, which is
a UV completion of General Relativity. In our quest we have used
gravitational collapse as the main tool. The focus is to probe the
nature of singularity (black hole or naked singularity) formed out
of the collapsing procedure. This is achieved via a geodesic
study. For our investigation we have considered two different
models of $f(R)$ gravity, namely the inflationary Starobinsky's
model and the power law model. Our study reveals the fact that
naked singularity is as good a possibility as black hole as far as
the central singularity is concerned. Via a proper fine tuning of
the initial data, we may realize both black hole or naked
singularity as the end state of the collapse. Thus this study is
extremely important and relevant in the light of the Cosmic
Censorship hypothesis. The most important result derived from the
study is that gravity's rainbow increases the tendency of
formation of naked singularities. We have also deduced the
conditions under which the singularity will be a strong or weak
curvature singularity. Finally in our quest to know more about the
model we have performed a thermodynamical study. Throughout the
study we have obtained results which involve deviation from the
classical set-up. Such deviations are expected in a quantum
evolution and can be attributed to the quantum fluctuations that
our model suffers from. It is expected that this study will
enhance our knowledge about quantization of gravity and
subsequently about the illusive theory of quantum gravity.
\end{abstract}

%%%%%%%%%%%%%%%%%%%%%%%%
\section{Introduction}
%%%%%%%%%%%%%%%%%%%%%%%%
More than a hundred years back, Einstein proposed his theory of
general relativity (GR) which revolutionized our idea of gravity.
It provided us a theory that could act as a powerful tool in our
quest to know the universe. With the passage of time and with
extensive research glitches in the theory started showing up. The
major blow came at the turn of the last century when the discovery
of the accelerated expansion of the universe \cite{acc1, acc2}
left GR inconsistent at cosmological distances. Since then we have
resorted to the alternative techniques of modified gravity and
dark energy to incorporate this accelerated expansion in our
theory of gravity. While the former deals with the geometry of
spacetime, the latter is concerned with the matter content of the
universe. Extensive reviews in modified gravity can be found in
the Refs.\cite{mod1, mod2, mod3}.

Many of such theories aim at modifying the linear function of
scalar curvature $R$ from its special form in GR to a more generic
form. $f(R)$ gravity is one such attempt where the gravitational
lagrangian of GR, $\mathcal{L}_{GR}=R$ is replaced by an analytic
function of $R$ i.e. $\mathcal{L}_{f(R)}=f(R)$. Choosing a
suitable function for $f(R)$, one can explore the non-linear
effects of the scalar curvature on the evolution of the universe.
Extensive reviews in $f(R)$ gravity can be found in the
Refs.\cite{fr1, fr2}. Viability of $f(R)$ dark energy models have
been studied in ref.\cite{frde}, where the $f(R)$ models with a
power law of $R$ has been ruled out. Author of
Ref.\cite{scalartensor} studied the interplay between $f(R)$
theories and scalar-tensor theories via the Palatini formalism.
Formation of large scale structure in $f(R)$ gravity was studied
in Ref.\cite{largescale}. A reconstruction scheme for $f(R)$
theories was explored in Ref.\cite{reconstruct}. Various other
studies related to $f(R)$ gravity can be found in Refs.\cite{fr3,
fr4, fr5, fr6}.

For a long time we have been searching for a theory of gravity
that will be consistent at all length scales and at all energy
levels. Such a theory is termed as the Theory of Everything (TOE).
It is understood that this would eventually boil down to a theory
of quantum gravity where GR will be reconciled to the theory of
quantum mechanics (QM). Till now there have been a few proposals
for such a theory namely, Loop quantum theory \cite{loop1, loop2},
String theory \cite{string1, string2}, Horava-Lifshitz (HL)
gravity \cite{hl1, hl2}, etc. The UV completion of GR in the limit
that GR is recovered in IR limit has eventually led to the
development of the Horava-Lifshitz gravity. An alternative
mechanism of UV completion of GR was proposed by Magueijo and
Smolin in the Ref.\cite{rainbow1} where the geometry of spacetime
depends on the energy of the test particle. This theory is termed
as Gravity's Rainbow. Although the conceptual basis of HL gravity
and gravity's rainbow are quite different yet they aim to pursue
similar ideas. Both the theories resorts to the modification of
the usual energy-momentum dispersion relations of the special
theory of relativity in the UV limit, such that in the IR limit
the usual relations are retrieved. Following this, the authors in
Ref.\cite{bridge} tried to bridge the two theories from the
conceptual background.

We know that in GR the usual energy-momentum relations are
governed by the Lorentz symmetry and so any modifications to these
in the UV limit will directly imply the violation of the symmetry.
In fact different quantum gravity approaches have shown that at
high energy scales (UV limit) Lorentz symmetry breaks down
\cite{uv1, uv2, uv3, uv4}. This breakdown is expected to occur in
models like string field theory \cite{stringfield}, discrete
spacetime \cite{discrete}, spacetime foam \cite{foam},
non-commutative geometry \cite{commutative}, spin network in Loop
quantum gravity \cite{slqg}, etc. Magueijo and Smolin in the
Ref.\cite{dsr} proposed the theory of Doubly Special Relativity
(DSR), where Einstein's Special Theory of Relativity (STR) is
generalized for spacetimes with high energy, i.e. energies
compared to Planck energy $(E_P=10^{19} GeV)$. In this theory
Planck energy joins the speed of light as an invariant quantity.
Nevertheless this formulation was achieved at the cost of the
violation of the Lorentz symmetry. The modifications to the field
theory also suggested corresponding modification to the
equivalence principle which determines how DSR can be embedded in
GR. Introducing non-zero curvature in DSR, we get what can be
called the Doubly General Relativity (DGR) or Gravity's Rainbow.
In this theory the modifications to the usual energy-momentum
dispersion relations $(E^{2}-p^{2}=m^{2})$ as discussed above is
introduced via energy dependent rainbow functions $\mathcal{F}(E)$
and $\mathcal{G}(E)$ as given below,
\begin{equation}\label{energymomentummod}
E^{2}\mathcal{F}^{2}(E)-p^{2}\mathcal{G}^{2}(E)=m^{2}
\end{equation}
Here $E=E_{s}/E_{P}$, where $E_{s}$ is the maximum energy that a
probe in the spacetime can support, and $E_P$ is the Planck
energy. From the conceptual background of the theory it is obvious
that $E_{s}$ cannot exceed $E_{P}$. The choice of the rainbow
functions should be such that they respect the correspondence
principle, i.e. at the IR limit we should be able to recover the
usual energy-momentum dispersion relations of classical GR from
the Eqn.(\ref{energymomentummod}). This means that the rainbow
functions are required to satisfy the relations,
\begin{equation}\label{correspondence}
\lim\limits_{E_s/E_P\to0} \mathcal{F}(E)=1,\qquad
\lim\limits_{E_s/E_P\to0} \mathcal{G}(E)=1.
\end{equation}
In the above relations the limit $E_S/E_P\to0$ corresponds to a
spacetime with low energy where GR becomes dominant. It is
expected that as $E_s\to E_P$, effects of GR gradually fades away
and quantum gravity effects become more and more dominant. The
rainbow functions are motivated from various theoretical and
phenomenological considerations in literature. Using the results
from loop quantum gravity and $\kappa$-Minkowski noncommutative
spacetime these functions have been proposed as \cite{mot1, mot2},
\begin{equation}\label{rainbowfuncchoice}
\mathcal{F}(E_s/E_P)=1, ~~~~ \mathcal{G}(E_s/E_P)=\sqrt{1-\eta
\frac{E_s}{E_P}}
\end{equation}
where $\eta$ is a constant. The modified dispersion from constant
velocity of light motivates the following rainbow functions
\cite{dsr}
\begin{equation}\label{rainbowfuncchoice1}
\mathcal{F}(E_s/E_P)=\mathcal{G}(E_s/E_P)=\frac{1}{1-a_{1}\frac{E_s}{E_P}}
\end{equation}
where $a_{1}$ is a constant. Moreover the observations of the hard
spectra from gamma ray bursters have been used motivate the
rainbow functions \cite{foam},
\begin{equation}\label{rainbowfuncchoice2}
\mathcal{F}(E_s/E_P)=\frac{e^{a_{2}E_s/E_P}-1}{a_{2}E_s/E_P}
,~~~~~~~\mathcal{G}(E_s/E_P)=1
\end{equation}
where $a_{2}$ is a constant. In this study it makes sense to use
the rainbow functions given by eqn.(\ref{rainbowfuncchoice})
because the choice is motivated from a theory of quantum gravity
(loop quantum gravity).

The metric in gravity's rainbow is written as
\begin{equation}\label{rainmetric}
g^{\mu\nu}(E)=\eta^{ab}e^\mu_a(E) e^\nu_b(E).
\end{equation}
where $e_0=\mathcal{F}^{-1}(E_s/E_P)\tilde{e_{0}}$ and
$e_i=\mathcal{G}^{-1}(E_s/E_P)\tilde{e_{i}}$. Here the tilde
quantities refer to the frame fields which are independent of
energy thus corresponding to the geometry explored by a low energy
quanta. Eqn.(\ref{rainmetric}) actually represents a family of
energy dependent metrics thus forming a rainbow. In such a
spacetime each test particle with different energy will probe a
different geometry thus following different geodesics. Due to its
quantum gravity background gravity's rainbow have been studied
extensively in recent times \cite{study1, study2, study3, study4,
study5, study6, study7, study8}.

Gravitational collapse is an astrophysical phenomenon that plays a
central role in the structure formation process of the universe.
For this reason gravitational collapse has been a field of
interest for astrophysicists over the years. It all started with
Oppenheimer and Snyder who studied the collapse of a dust cloud
with a static Schwarzschild exterior and Friedmann like interior
\cite{oppen}. Subsequently the collapse of spherically symmetric
inhomogeneous distribution of dust was studied by Tolman
\cite{tolman} and Bondi \cite{bondi}. In 1969, Roger Penrose in a
phenomenal paper \cite{penrose} argued that any types of
cosmological singularity is bound to be shrouded by an event
horizon thus making it a Black Hole (BH). This proposal is
popularly known in literature as the Cosmic Censorship Hypothesis
(CCH). But this idea has subsequently been questioned in the
absence of any form of rigorous proof or observational evidence.
This naturally led people to search for collapsing models that can
yield singularities which are uncensored, commonly known as Naked
Singularities (NS) \cite{ns1, ns2, ns3, ns4, ns5, cch2, cch3}.
Such an uncensored singularity will be really interesting to study
because it will leak information that is generally hidden behind
the event horizon of a BH. Not only the solution of information
paradox but a proper knowledge of NS will eventually help us
understand the gravity quanta thus allowing us to formulate the
illusive satisfactory theory of quantum gravity.

In 1951, P.C. Vaidya formulated a relativistic line element
representing the field of radiation for a non-static mass
\cite{v1}. This is a generalization of the Schwarzschild solution
for non-static mass. Schwarzschild's external solution can
represent the gravitational field of a cold dark body with a
constant mass. So it is obvious that the application of this
solution to describe the sun's gravitational field should only be
considered as approximate. Vaidya's metric precisely solved this
problem by successfully describing the spacetime of Sun. In fact
it can represent the spacetime of any radiating mass, such as a
star. This is why it is sometimes called the {\it radiating or
shining Schwarzschild metric}. It should be mentioned here that
this metric, if expressed in radiation coordinates, differs from
the Schwarzschild metric only in that the constant mass parameter
$m$ is replaced by a function of retarded time. Notable studies
using gravity's rainbow in Vaidya's metric can be found in
\cite{s1, s2, s3}. Other important studies on Vaidya spacetime can
be found in \cite{s4, s5, s6, s7, s8}.

From the above discussion we feel the need to explore the
non-static radiating Vaidya spacetime in a quantum regime. This
can be achieved by introducing rainbow deformations in the Vaidya
spacetime in the background of a gravity theory. $f(R)$ gravity
being the simplest and most obvious theory of modified gravity at
least at the mathematical level, we are inclined to consider it in
the background of our model. It is expected that the radiating
star represented by Vaidya spacetime will yield very interesting
results in the quantum limit. The model will be investigated via a
study of gravitational collapse, which is a very important
astrophysical phenomenon. The study will be eventually
complimented by a thermodynamical study in the said model. The
search for the illusive theory of quantum gravity is a motivation
for the present work and we hope that our investigation will be a
step towards the better understanding of the nature of gravity
quanta. The paper is organized as follows. In Sec.2, the field
equations for the rainbow deformed Vaidya spacetime in $f(R)$
gravity are derived and a solution is obtained. Sec.3 is dedicated
to the study of gravitational collapse in the system. In Sec.4 we
have studied the thermodynamical properties of the system in
detail. Finally the paper is concluded with a conclusion and
discussion in Sec.5.

%%%%%%%%%%%%%%%%%%%%%%%%%%%%%%%%%%%%%%%%%%%%%%%%%%%%%%%%%%%%%
\section{Rainbow deformed Vaidya spacetime in $f(R)$ gravity}
%%%%%%%%%%%%%%%%%%%%%%%%%%%%%%%%%%%%%%%%%%%%%%%%%%%%%%%%%%%%%

The Einstein-Hilbert action of GR is given by,
\begin{equation}\label{actionEH}
S_{EH}=\frac{1}{2\kappa}\int d^{4}x\sqrt{-g}R
\end{equation}
where $\kappa\equiv 8\pi G$, $G$ is the gravitational constant,
$g$ is the determinant of the metric and $R$ is the Ricci scalar
(we have considered $c=1$). We replace the Ricci scalar, $R$ in
the above action by a generalized function of $R$ to get the
action for $f(R)$ gravity,
\begin{equation}\label{action}
S=\frac{1}{2\kappa}\int d^{4}x\sqrt{-g}f(R)
\end{equation}
Beginning from the action (\ref{action}) and adding a matter term
$S_M$, the total action for $f(R)$ gravity takes the form,
\begin{equation}\label{actiontotal}
S=\frac{1}{2\kappa}\int d^{4}x\sqrt{-g}f(R)+S_M(g_{\mu\nu}, \psi)
\end{equation}
where $\psi$ collectively denotes the matter fields. Taking
variation with respect to the metric we get the field equations
as,
\begin{equation}\label{field}
f'(R)R_{\mu\nu}-\frac{1}{2}g_{\mu\nu}f(R)+\left(g_{\mu\nu}\Box-\nabla_{\mu}\nabla_{\nu}\right)f'(R)=\kappa
T_{\mu\nu}
\end{equation}
where $T_{\mu\nu}$ is given by,
\begin{equation}\label{stressenergy}
T_{\mu\nu}=-\frac{2}{\sqrt{-g}}\frac{\delta S_M}{\delta
g^{\mu\nu}}
\end{equation}
In the above equations a prime denotes differentiation with
respect to the argument, i.e. $R$. $\nabla_{\mu}$ denotes
covariant derivative associated with the Levi-Civita connection of
the metric and $\Box\equiv \nabla^{\mu}\nabla_{\mu}$ is the
D'Alembertian operator. The field equations given by equation
(\ref{field}) can also be written in the following form,
\begin{equation}\label{fieldeqn}
G_{\mu\nu}=R_{\mu\nu}-\frac{1}{2}g_{\mu\nu}R=\frac{\kappa
T_{\mu\nu}}{f'(R)}+g_{\mu\nu}\frac{\left[f(R)-R f'(R)\right]}{2
f'(R)} +\frac{\left[\nabla_{\mu}\nabla_{\nu}f'(R)-g_{\mu\nu}\Box
f'(R)\right]}{f'(R)}
\end{equation}

The Vaidya metric in the advanced time coordinate system is given
by,
\begin{equation}\label{vaidya}
ds^{2}=f(t,r)dt^{2}+2dtdr+r^{2}\left(d\theta^{2}+\sin^{2}\theta
d\phi^{2}\right)
\end{equation}
~~~~~where $f(t,r)=-\left(1-\frac{m(t,r)}{r}\right)$. It is
obvious that here we have used the units $G=c=1$.

Introducing rainbow deformations in the above metric we
get\cite{rainbow1},
\begin{equation}\label{vaidyarainbow}
ds^{2}=-\frac{1}{\mathcal{F}^2(E)}\left(1-\frac{m(t,r)}{r}\right)dt^{2}+\frac{2}{\mathcal{F}(E)\mathcal{G}(E)}dtdr+\frac{r^{2}}{\mathcal{G}^2(E)}\left(d\theta^{2}+\sin^{2}\theta
d\phi^{2}\right)
\end{equation}
The total energy momentum tensor of the field equation
(\ref{fieldeqn}) is given by the following sum,
\begin{equation}\label{energymom}
T_{\mu\nu}=T_{\mu\nu}^{(n)}+T_{\mu\nu}^{(m)}
\end{equation}
where $T_{\mu\nu}^{(n)}$ and $T_{\mu\nu}^{(m)}$ are the
contributions from the Vaidya null radiation and perfect fluid
respectively defined as,
\begin{equation}\label{null}
T_{\mu\nu}^{(n)}=\sigma l_{\mu}l_{\nu}
\end{equation}
and
\begin{equation}\label{fluid}
T_{\mu\nu}^{(m)}=(\rho+p)(l_{\mu}\eta_{\nu}+l_{\nu}\eta_{\mu})+pg_{\mu\nu}
\end{equation}
where $\rho$ and $p$ are the energy density and pressure for the
perfect fluid and $\sigma$ is the energy density corresponding to
Vaidya null radiation. In the co-moving co-ordinates
($t,r,\theta_{1},\theta_{2},...,\theta_{n}$), the two eigen
vectors of energy-momentum tensor namely $l_{\mu}$ and
$\eta_{\mu}$ are linearly independent future pointing null vectors
having components
\begin{equation}\label{vectors1}
l_{\mu}=(1,0,0,0)~~~~ and~~~~
\eta_{\mu}=\left(\frac{1}{2}\left(1-\frac{m}{r}\right),-1,0,0\right)
\end{equation}
and they satisfy the relations
\begin{equation}\label{vectors2}
l_{\lambda}l^{\lambda}=\eta_{\lambda}\eta^{\lambda}=0,~
l_{\lambda}\eta^{\lambda}=-1
\end{equation}
Now we impose rainbow deformations on the linearly independent
future pointing null vectors $l_{\mu}$ and $\eta_{\mu}$ such that
we get,
\begin{equation}\label{rainbowvectors1}
l_{\mu}=\left(\frac{1}{\mathcal{F}(E)},0,0,0\right)~~~~~\&~~~
\eta_{\mu}=\left(\frac{1}{2\mathcal{F}(E)}\left(1-\frac{m(t,r)}{r}
\right),-\frac{1}{\mathcal{G}(E)},0,0  \right)
\end{equation}
satisfying the following conditions (\ref{vectors2}). Therefore,
the non-vanishing components of the total energy-momentum tensor
will be as follows
\begin{eqnarray*}
T_{00}=\frac{\sigma}{\mathcal{F}^{2}(E)}+\frac{\rho}{\mathcal{F}^{2}(E)}\left(1-\frac{m(t,r)}{r}\right),&&
~~T_{01}=-\frac{\rho}{\mathcal{F}(E)\mathcal{G}(E)}, \\
\end{eqnarray*}
\begin{eqnarray}\label{energymomentum}
T_{22}=\frac{pr^2}{\mathcal{G}^2(E)}, && ~~T_{33}=\frac{pr^2
\sin^2\theta}{\mathcal{G}^2(E)}
\end{eqnarray}
Here we consider matter in the form of perfect barotropic fluid
given by the equation of state
\begin{equation}
p=\omega\rho
\end{equation}
where '$\omega$' is the barotropic parameter.

The non-vanishing components of the Einstein tensors are given by,
\begin{eqnarray*}
G_{00}=\frac{\mathcal{G}(E)}{r^{3}\mathcal{F}^{2}(E)}\left\{\mathcal{G}(E)\left(r-m\right)m'+\mathcal{F}(E)r
\dot{m}\right\},&&
~~~~G_{01}=G_{10}=-\frac{\mathcal{G}(E)m'}{r^{2}\mathcal{F}(E)} \\
\end{eqnarray*}
\begin{eqnarray}\label{einsteintensors}
G_{22}=-\frac{1}{2}r m'', && ~~G_{33}=-\frac{1}{2}r m''~
\sin^{2}\theta
\end{eqnarray}
where dot and dash represents derivatives with respect to time $t$
and radial coordinate $r$ respectively. For this system the Ricci
scalar becomes,
\begin{equation}\label{ricciscalar}
R=\frac{\mathcal{G}^{2}(E)}{r^{2}}\left(2m'+rm''\right)
\end{equation}

\vspace{4mm}

\subsection{Field Equations}

Here we report the computed field equation for $f(R)$ gravity's
rainbow in the time dependent Vaidya spacetime. In the study that
follows we have made use of the $(11)$ component of the field
equations because of its simplicity and so we report only that
component in this section. The other components of the field
equations are reported in the appendix section.\\

\textbf{\textit{The (11)-component of field equations is given
by,}}
\begin{equation}\label{11}
f''(R)r^{2}\left(12m'-6m''r+m^{iv}r^{3}\right)+f'''(R)\mathcal{G}^{2}(E)\left\{-4m'+r\left(m''+m'''r\right)\right\}^{2}=0
\end{equation}
where $m^{iv}$ denotes the fourth partial derivative of $m$ with
respect to $r$.

\subsection{Solution of the system} To find a solution of the field
equations we have to consider specific models of $f(R)$ gravity.
Here we consider two popular models, namely the inflationary
Starobinsky's model and the power law model, and subsequently find
the solution of the system. The emergence of the cosmic structure
from the homogeneous and isotropic universe cannot be clearly
explained by the standard inflationary models, because the
mechanism involved preserves the homogeneity and isotropy at all
times. A solution to this problem has been proposed by Sudarsky et
al. in Refs.\cite{infla1, infla2, infla3, infla4, infla5, infla6,
infla7, infla8, infla9}, where they have introduced the concept of
self induced collapse hypothesis. Here the collapse of the wave
function of the inflaton mode is restricted to occur during the
inflationary period. So there are reasons to believe that
collapsing scenario in inflationary models can help us understand
the quantum evolution of universe. Moreover quantum corrections
induced by rainbow functions will be highly compatible with such a
model. This motivates us to choose the Starobinsky's model for our
study. The motivation for the power law model is obvious as it is
the most generic model of $f(R)$ gravity capable of representing
all the epochs of the universe by fine tuning the initial data.

\subsubsection{Starobinsky's Model}
In this section we consider the popular Starobinsky's inflationary
model of $f(R)$ gravity. The model is given as \cite{staro1,
staro2}

\begin{equation}\label{star}
f(R)=R+aR^2
\end{equation}
where $a\geq 0$ is the only free parameter of the theory and has
the dimensions of $[mass]^{-2}$. So the parameter $a$ can also be
written in the form $a=1/M^{2}$, where mass becomes the free
parameter of the gravity theory. It can easily be seen that GR can
be retrieved from the theory for $a=0$. Using eqns.
(\ref{ricciscalar}),(\ref{11}) and (\ref{star}) we get the
following differential equation in terms of the mass '$m$',
\begin{equation}
2ar^{2}\left(r^{3}m^{iv}-6rm''+12m'\right)=0
\end{equation}
We can see that for $a=0$, we get an identity which corresponds to
the case for GR. From the above equation we get,
\begin{equation}\label{massdiff2}
r^{3}m^{iv}-6rm''+12m'=0, ~ ~~a\neq 0
\end{equation}
It should be noted that the parameter $'a'$ does not appear in the
above differential equation. Solving the above differential
equation we get,
\begin{equation}\label{mass2}
m(t,r)=f_{1}(t)-\frac{f_{2}(t)}{r}+f_{3}(t)r^{3}+f_{4}(t)r^{4}
\end{equation}
where $f_{1}(t)$, $f_{2}(t)$, $f_{3}(t)$ and $f_{4}(t)$ are
arbitrary functions of time $t$. So for the Starobinsky's model,
the rainbow deformed Vaidya spacetime in $f(R)$ gravity is given
by,
\begin{equation}\label{rainbowstaro}
ds^{2}=-\frac{1}{\mathcal{F}^2(E)}\left[1-\frac{1}{r}\left(f_{1}(t)-\frac{f_{2}(t)}{r}+f_{3}(t)r^{3}+f_{4}(t)r^{4}\right)\right]dt^{2}+\frac{2}{\mathcal{F}(E)\mathcal{G}(E)}dtdr+\frac{r^{2}}{\mathcal{G}^2(E)}\left(d\theta^{2}+\sin^{2}\theta
d\phi^{2}\right)
\end{equation}

\subsubsection{Power Law Model}
Regarding power law model we would like to state that probably it
is the most generic model for any theory and has a wide range of
applicability. Hence it is not surprising that it will reduce to
Starobinsky's model for some physical restrictions. But it should
be mentioned here that one should be cautious in using a generic
form of power law model because for all values of the exponent the
model may not remain cosmologically viable or for all values the
model may not satisfy the standard observational tests. Amendola
et al. \cite{frde} points out such a discrepancy and rules out
some basic models of f(R) theory. There are other examples as
well. In the light of these results any unhindered and free use of
the power law model will be worrying. However here our concern is
to keep our study as generic as possible and not bother about
observational compliances. Since our basic aim is to study
gravitational collapse which is an astrophysical phenomenon we
have restricted our study to just astrophysical implications.

We consider the following power law model \cite{rev1, rev2} of
$f(R)$ gravity,
\begin{equation}\label{pow}
f(R)=\alpha R^n~, ~~~~~n>0
\end{equation}
where $\alpha$ and $n>0$ are constants. Using
eqns.(\ref{ricciscalar}),(\ref{11}) and (\ref{pow}) we get the
following differential equation in terms of the mass '$m$',
\begin{eqnarray*}
\mathcal{G}^{2}(E)\alpha~
n\left(n-1\right)\left[\frac{\mathcal{G}^{2}(E)}{r^{2}}\left(rm''+2m'\right)\right]^{n-3}\left[\left(n-2\right)
\left(r^{2}m'''+rm''-4m'\right)^{2}\right.
\end{eqnarray*}
\begin{equation}\label{massdiff1}
\left.+\left(rm''+2m'\right)\left(r^{3}m^{iv}-6rm''+12m'\right)\right]=0
\end{equation}
It is almost impossible to get a general solution for the above
equation by the known mathematical methods. So we may discuss some
particular cases. It can easily be seen that the equation becomes
an identity for $n=0$ and $n=1$ and does not possess a viable
solution. Moreover these are trivial cases for the power law and
are of very little interest. For $n=2$ this equation reduces to
the corresponding equation for the Starobinsky model discussed in
the previous section.

The equation can be separated into two component equations as
given below:
\begin{equation}\label{comp1}
rm''+2m'=0, ~~~~~for~~ n\geq 4
\end{equation}
OR
\begin{equation}\label{comp2}
\left(n-2\right)\left(r^{2}m'''+rm''-4m'\right)^{2}+\left(rm''+2m'\right)\left(r^{3}m^{iv}-6rm''+12m'\right)=0
\end{equation}
For eqn.(\ref{comp1}) the solution is obtained as,
\begin{equation}\label{soln1}
m(t,r)=f_{5}(t)-\frac{f_{6}(t)}{r}
\end{equation}
where $f_{5}(t)$ and $f_{6}(t)$ are arbitrary functions of time,
$t$. Eqn.(\ref{comp2}) gives two alternative solutions only for
$n=2$. The solutions are given by,
\begin{equation}\label{soln2}
m(t,r)=f_{7}(t)-\frac{f_{8}(t)}{r}
\end{equation}
OR
\begin{equation}\label{soln3}
m(t,r)=f_{9}(t)-\frac{f_{10}(t)}{r}+f_{11}(t)r^{3}+f_{12}(t)r^{4}
\end{equation}
where $f_{i}(t),~ i=7, 8, 9, 10, 11, 12$ are arbitrary functions
of time. Combining eqns.(\ref{soln2}) and (\ref{soln3}) we get
(for $n=2$),
\begin{equation}\label{soln4}
m(t,r)=A\left(f_{7}(t)-\frac{f_{8}(t)}{r}\right)+B\left(f_{9}(t)-\frac{f_{10}(t)}{r}+f_{11}(t)r^{3}+f_{12}(t)r^{4}\right)
\end{equation}
where $A$ and $B$ are arbitrary constants. Since solution
(\ref{soln1}) and (\ref{soln4}) are valid for different values of
$n$, they cannot be combined to get a single expression for
$m(t,r)$. So the expression for $m(t,r)$ becomes,
\begin{equation}\label{massfunc1}
m(t,r)=\left\{\begin{array}{c}
 f_{5}(t)-\frac{f_{6}(t)}{r},~~~~when~n\geq 4\\\\
A\left(f_{7}(t)-\frac{f_{8}(t)}{r}\right)+B\left(f_{9}(t)-\frac{f_{10}(t)}{r}+f_{11}(t)r^{3}+f_{12}(t)r^{4}\right),~~~~when~n=2
\end{array}\right.
\end{equation}
Hence for the power law model, the rainbow deformed Vaidya
spacetime in $f(R)$ gravity is given by,
\begin{equation}\label{vaidyapower}
ds^{2}=\left\{\begin{array}{c}
-\frac{1}{\mathcal{F}^2(E)}\left[1-\frac{1}{r}\left(f_{5}(t)-\frac{f_{6}(t)}{r}\right)\right]dt^{2}+\frac{2}{\mathcal{F}(E)\mathcal{G}(E)}dtdr+\frac{r^{2}}{\mathcal{G}^2(E)}\left(d\theta^{2}+\sin^{2}\theta
d\phi^{2}\right),\\\\
~~~~when~n\geq 4\\\\
-\frac{1}{\mathcal{F}^2(E)}\left[1-\frac{1}{r}\left\{A\left(f_{7}(t)-\frac{f_{8}(t)}{r}\right)+B\left(f_{9}(t)-\frac{f_{10}(t)}{r}+f_{11}(t)r^{3}+f_{12}(t)r^{4}\right)\right\}\right]dt^{2}+\frac{2}{\mathcal{F}(E)\mathcal{G}(E)}dtdr\\\\
+\frac{r^{2}}{\mathcal{G}^2(E)}\left(d\theta^{2}+\sin^{2}\theta
d\phi^{2}\right),~~~~when~n=2
\end{array}\right.
\end{equation}

%\subsubsection{Logarithmic Model}
%The Logarithmic model of $f(R)$ gravity \cite{log1, log2} is given
%by
%\begin{equation}\label{log}
%f(R)=ln\left(\lambda R\right)
%\end{equation}

%\subsubsection{Exponential Model}
%The exponential model of $f(R)$ gravity \cite{exp1, exp2, exp3,
%exp4} is given by
%\begin{equation}\label{exp}
%f(R)=exp\left(\beta R\right)
%\end{equation}
%where $\beta$ is an arbitrary real number.

\section{Gravitational Collapse}
In this section we aim to study the phenomenon of gravitational
collapse of a star modelled by the deformed Vaidya metric given in
eqn.(\ref{vaidyarainbow}). We will use the solutions obtained in
eqns.(\ref{rainbowstaro}) and (\ref{vaidyapower}) for the two
different models separately. The methodology to be employed to
study the collapsing procedure is the geodesic study. We will be
interested in probing the existence of non-spacelike radial
geodesics emanating from the central singularity, that were
terminated in the past at the singularity $r=0$. A geodesic coming
out from the central singularity and interacting with the outside
would imply the non existence of the event horizon around the
singularity. Such a singularity is quite aptly termed as a naked
singularity. Due to the absence of any form of horizon an outside
observer can receive information from such a singularity and
vice-versa. If only a single null geodesic escapes from the
singularity, it indicates the emission of a single wavefront and
hence the singularity would be visible (naked) only momentarily to
a distant observer. This singularity is locally naked. On the
other hand if the NS is to be visible for a finitely prolonged
time period, a family of geodesics must leave the central
singularity, which will make the singularity globally naked. If
our search for the emanating geodesics yield negative result, then
the singularity is bound to be a BH and we will get yet another
reason to upheld the cosmic censorship hypothesis, which does not
have a rigorous proof till date. On the contrary there are quite a
few works in literature \cite{ns1, cch2, cch3} that supports the
formation of NS thus providing significant counterexamples for the
CCH as discussed earlier. In spite of these counterexamples,
generically in classical background, CCH holds good and a
singularity is always censored but in this work it is expected
that the quantum nature of gravity will play its role and support
the formation of NS.

Considering the collapse to be spherical, we assume the physical
radius of the $r$-th shell of the star at time $t$ be $R(t,r)$. In
the epoch $t=0$, we have $R(0,r)=r$. If the collapse is
inhomogeneous, then different shells may become singular at
different times. If the outgoing non-spacelike geodesics possess
well defined tangent at the singularity, $\frac{dR}{dr}$ will tend
to a finite limit as the geodesic approaches the singularity in
the past along the trajectories. As the trajectories reach the
points $(t_{0}, r)=(t_{0}, 0)$, the singularity occurs, which is
given by $R(t_{0}, 0)=0$. Physically this corresponds to the
matter shells being crushed to zero radius, resulting in the
formation of the central singularity. If we trace back the
trajectories of the outgoing non-spacelike geodesics from this
central singularity, it is likely that they will terminate in the
past at the singularity $(r=0, t=t_0)$ where $R(t_{0},0)=0$.
Therefore we should have $R\rightarrow 0$ as $r\rightarrow 0$
\cite{sing}.

The equation for outgoing radial null geodesics can be obtained
from equation (\ref{vaidyarainbow}) by putting $ds^{2}=0$ and
$d\Omega_{2}^{2}=d\theta^{2}+\sin^{2}\theta d\phi^{2}=0$ as
\begin{equation}
\frac{dt}{dr}=\frac{2\mathcal{F}(E)}{\mathcal{G}(E)\left(1-\frac{m(t,r)}{r}\right)}.
\end{equation}
From the above expression it is quite clear that at $r=0,~t=0$
there is a singularity of the above differential equation. Suppose
we consider a parameter $X=\frac{t}{r}$. Using this parameter we
can study the limiting behaviour of the function $X$ as we
approach the singularity at $r=0,~t=0$ along the radial null
geodesic. If we denote the limiting value by $X_{0}$ then using
L'Hospital's rule we have
\begin{eqnarray}\label{X0}
\begin{array}{c}
X_{0}\\\\
{}
\end{array}
\begin{array}{c}
=\lim X \\
\begin{tiny}t\rightarrow 0\end{tiny}\\
\begin{tiny}r\rightarrow 0\end{tiny}
\end{array}
\begin{array}{c}
=\lim \frac{t}{r} \\
\begin{tiny}t\rightarrow 0\end{tiny}\\
\begin{tiny}r\rightarrow 0\end{tiny}
\end{array}
\begin{array}{c}
=\lim \frac{dt}{dr} \\
\begin{tiny}t\rightarrow 0\end{tiny}\\
\begin{tiny}r\rightarrow 0\end{tiny}
\end{array}
\begin{array}{c}
=\lim \frac{2\mathcal{F}(E)}{\mathcal{G}(E)\left(1-\frac{m(t,r)}{r}\right)} \\
\begin{tiny}t\rightarrow 0\end{tiny}~~~~~~~~~~~~\\
\begin{tiny}r\rightarrow 0\end{tiny}~~~~~~~~~~~~
 {}
\end{array}
\end{eqnarray}
This would indeed give an algebraic equation in terms of $X_{0}$.
Now any positive real root of this equation will give the
direction of the tangent to an outgoing null geodesic at the
singularity. So the existence of positive real roots to this
equation is a necessary and sufficient condition for the
singularity to be naked. Now as discussed earlier, if one single
null geodesic in the $(t,r)$ plane escapes the singularity, it
would mean that a single wavefront is emitted from the singularity
and hence the singularity would appear to be naked only
instantaneously to a distant observer. This would result in a
locally naked singularity. If the singularity is to be seen for a
finite period of time a family of null geodesics must escape from
the singularity thus making it globally naked. These can be
investigated from the number of real positive roots obtained from
the above equation. Now we will consider the results for the two
models separately.

\subsection{Starobinsky's Model}
 Using equations
(\ref{mass2}) and (\ref{X0}), we have
\begin{eqnarray}\label{X01}
\frac{2}{X_{0}}=
\begin{array}llim\\
\begin{tiny}t\rightarrow 0\end{tiny}\\
\begin{tiny}r\rightarrow 0\end{tiny}
\end{array}
\frac{\mathcal{G}(E)}{\mathcal{F}(E)}\left[1-\frac{f_{1}(t)}{r}+\frac{f_{2}(t)}{r^2}-f_{3}(t)r^{2}-f_{4}(t)r^{3}\right]
\end{eqnarray}
The functions $f_{i}(t)$, $t=1, 2, 3, 4$  are in fact constants of
integration that we get on solving (integrating) the field
equations. So in principle they should be obtained from suitable
initial or boundary conditions and physically should represent the
mass of the system. Obviously these initial and boundary
conditions will correspond to some particular physically viable
cases of the generic mathematical solutions. Therefore we should
consider those particular cases which are compatible with the
present study of gravitational collapse. Our choice for these
functions should be guided by the fact that for some suitable
transformations we should be able to retrieve those chosen
functions as initial or boundary conditions for our system. We
further mention that we will only consider self similar
expressions for these functions to suit our analysis. Non self
similar expressions can also be considered with consequences.
We choose the following: \\\\
~~~~~~~$f_{1}(t)=\gamma_{1} t$, ~~~~$f_{2}(t)=\delta_{1} t^{2}$,
~~~~$f_{3}(t)=\xi_{1} t^{-2}$, ~~~~$f_{4}(t)=\epsilon_{1} t^{-3}$
\\\\
where $\gamma_{1}$, $\delta_{1}$, $\xi_{1}$ and $\epsilon_{1}$ are
arbitrary constants. It should be clear that the above choices
have been made depending on the definition of $X_{0}$ in equation
(\ref{X0}) such that the ratio $t/r$ can be formed and
correspondingly its limit can be evaluated. So we have actually
considered such values of the arbitrary functions of time so that
the limit given by eqn.(\ref{X01}) exists.

Using the above chosen functions in eqn.(\ref{X01}) we get the
following algebraic equation in $X_{0}$
\begin{equation}\label{algebra1}
\delta_{1} X_{0}^{5}-\gamma_{1}
X_{0}^{4}+X_{0}^{3}-\frac{2\mathcal{F}(E)}{\mathcal{G}(E)}X_{0}^{2}-\xi_{1}
X_{0}-\epsilon_{1}=0
\end{equation}
The above equation being a five degree equation and hence it is
highly unlikely to find the solution and get the roots by the
known mathematical methods. Hence we try to plot the $X_{0}$
against the other parameters and get an idea about its trend. Such
plots have been generated in Figs.1, 2, 3 and 4. Also in fig.5 we
have shown the effect of gravity's rainbow on the collapsing
scenario and compared the results with those of general
relativity.

\begin{figure}
~~~~~~~~\includegraphics[height=1.7in,width=2.2in]{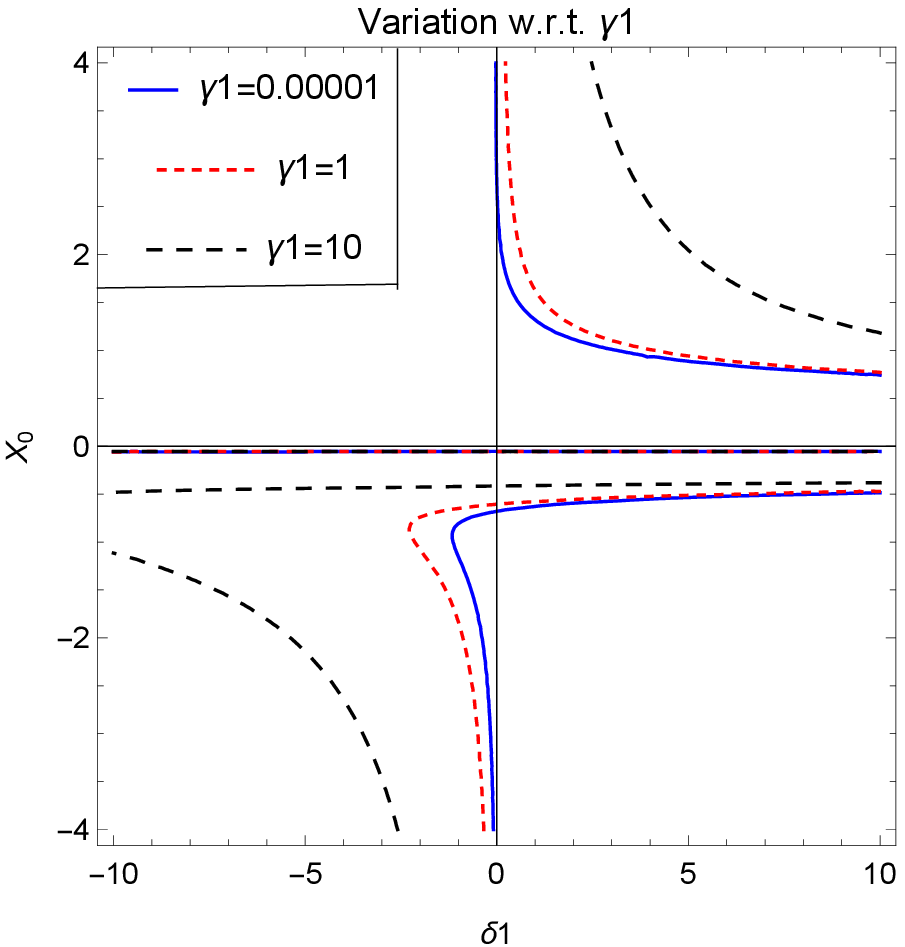}~~~~~~~~~~~~~\includegraphics[height=1.7in,width=2.2in]{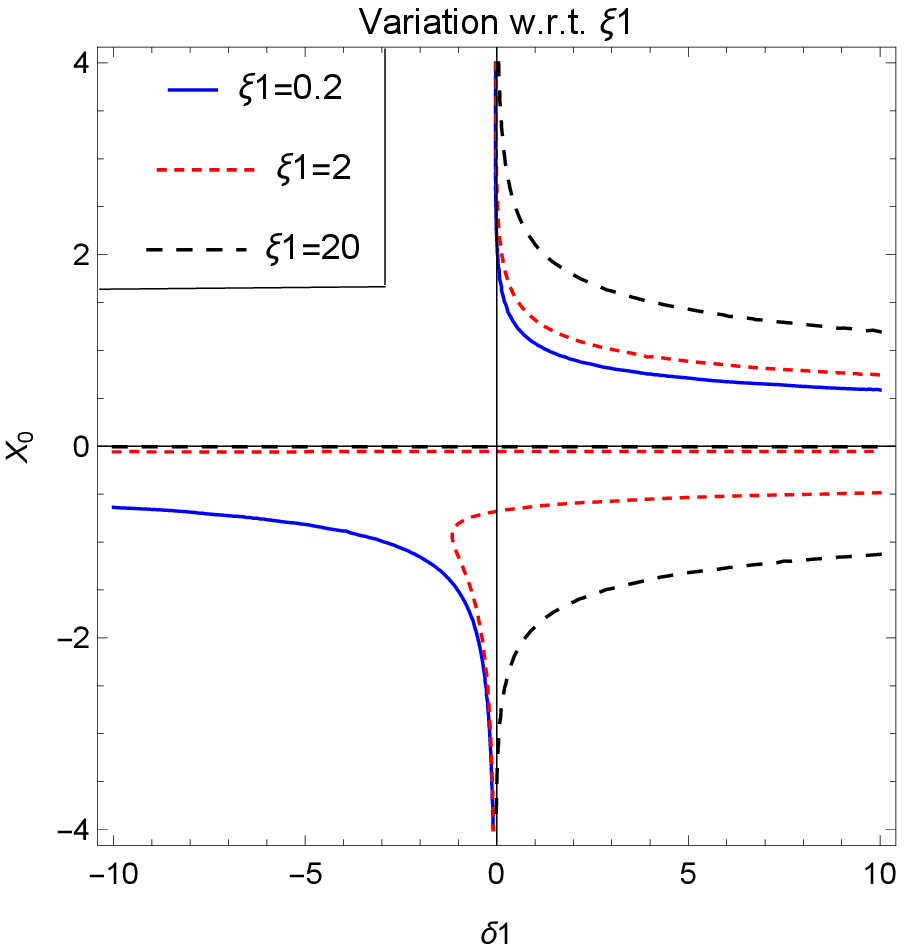}~~~~~~~\\

~~~~~~~~~~~~~~~~~~~~~~~~~~~~Fig.1~~~~~~~~~~~~~~~~~~~~~~~~~~~~~~~~~~~~~~~~~~~~~~~~~~~~~~~~~~Fig.2~~~~~~~~~\\

~~~~~~~~\includegraphics[height=1.7in,width=2.2in]{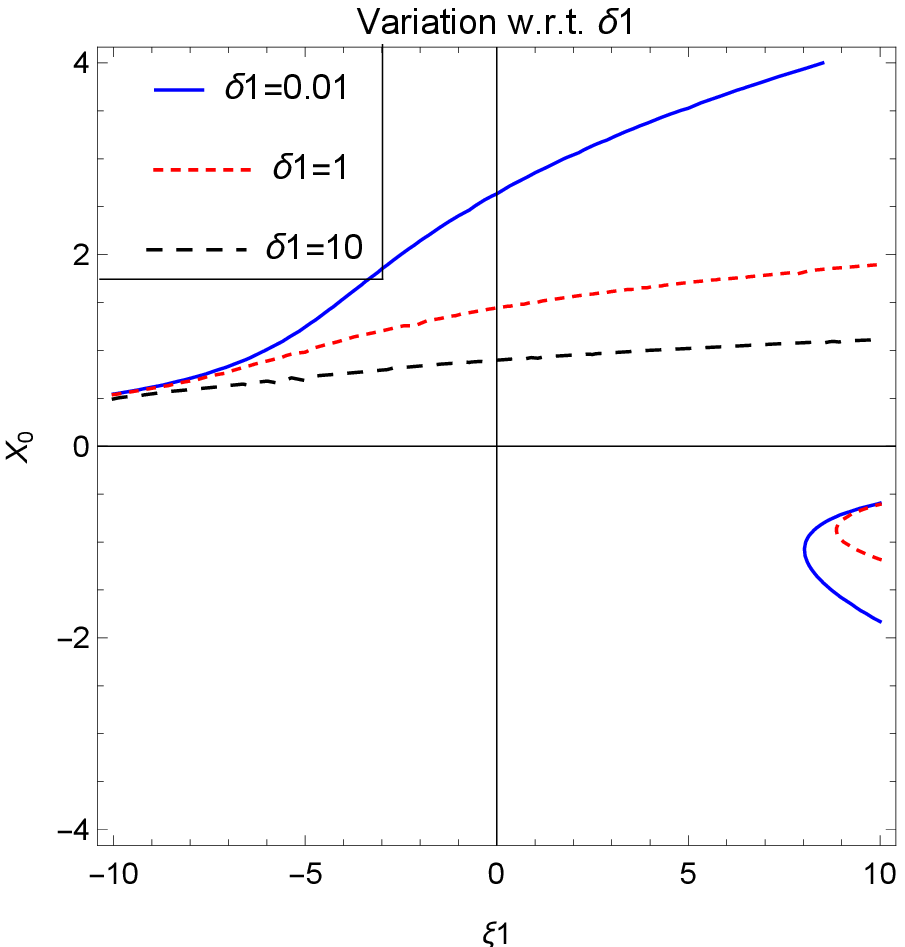}~~~~~~~~~~~~~\includegraphics[height=1.7in,width=2.2in]{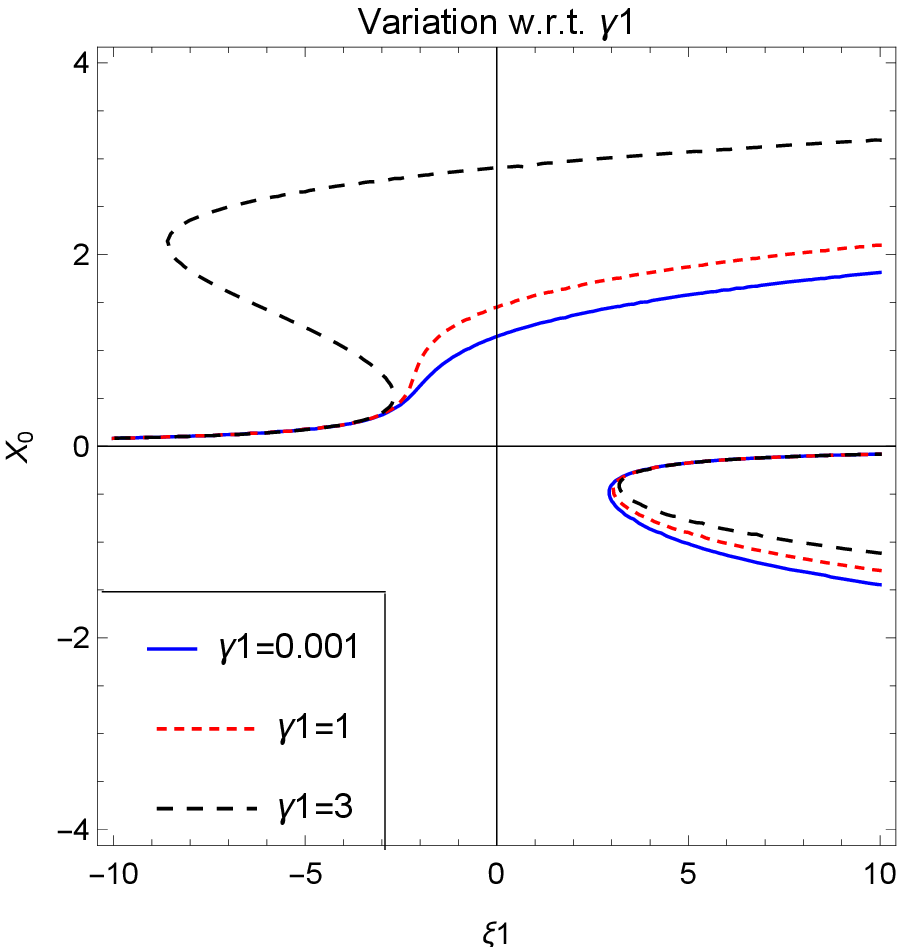}~~~~~~~\\

~~~~~~~~~~~~~~~~~~~~~~~~~~~~Fig.3~~~~~~~~~~~~~~~~~~~~~~~~~~~~~~~~~~~~~~~~~~~~~~~~~~~~~~~~~~~~Fig.4~~~~~~~~~\\

\vspace{1mm} \textit{\textbf{Figs.1 and 2} show the variation of
$X_{0}$ with $\delta_{1}$ for different values of $\gamma_1$ and
$\xi_1$ respectively for the Starobinsky's model. The Rainbow
functions are taken as $\mathcal{F}(E)=1$,
$\mathcal{G}(E)=\sqrt{1-\eta
\left(\frac{E_{s}}{E_{P}}\right)^{n}}$. In Fig.1 the initial
conditions are taken as $\xi_1=2$, $\epsilon_1=0.1$, $\eta=1$,
$n=2$, $E_{s}=1$ and $E_{P}=5$. In Fig.2 the initial conditions
are $\gamma_1=0.00001$, $\epsilon_1=0.1$, $\eta=1$, $n=2$,
$E_{s}=1$ and $E_{P}=5$.}\\\\
\vspace{1mm} \textit{\textbf{Figs.3 and 4} show the variation of
$X_{0}$ with $\xi_{1}$ for different values of $\delta_1$ and
$\gamma_1$ respectively for the Starobinsky's model. In Fig.3 the
initial conditions are taken as $\gamma_1=0.01$, $\epsilon_1=5$,
$\eta=1$, $n=2$, $E_{s}=1$ and $E_{P}=5$. In Fig.4 the initial
conditions are $\delta_1=1$, $\epsilon_1=0.8$, $\eta=1$, $n=2$,
$E_{s}=1$ and $E_{P}=5$. The rainbow functions are same as that of
the previous figures.}
\end{figure}

\begin{figure}
~~~~~~~~~~~~~~~~~~~~~~~~~~\includegraphics[height=2.5in,width=2.5in]{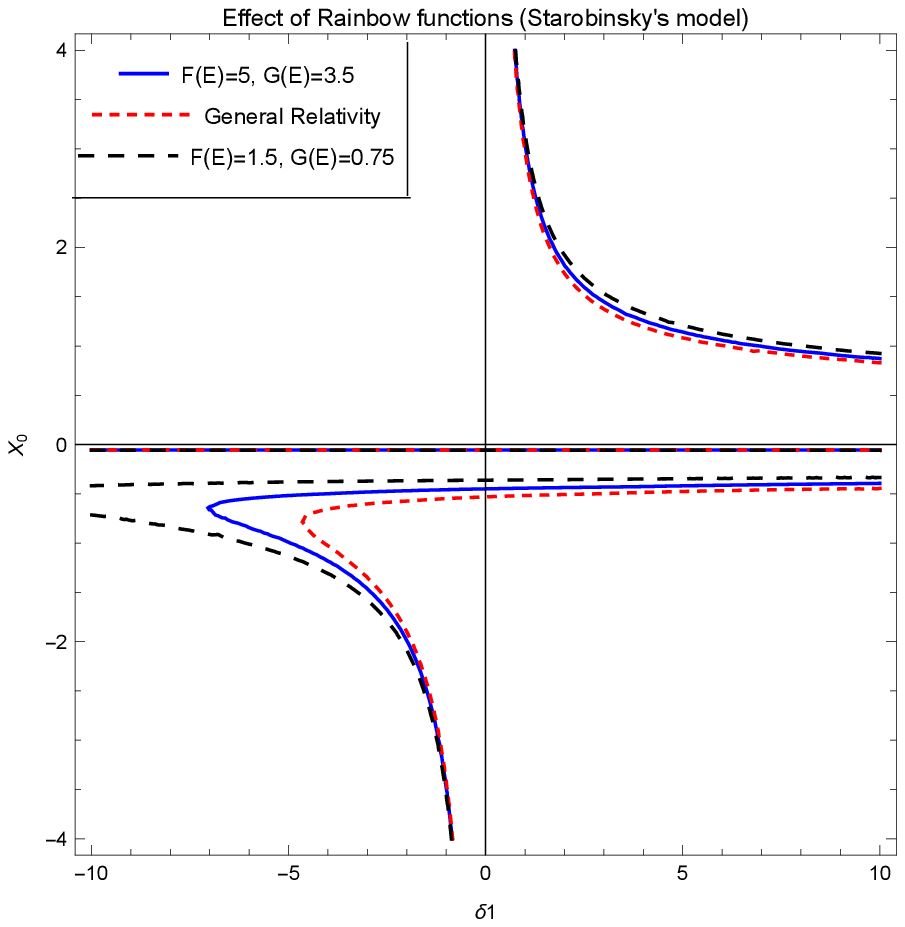}~~~~~~~~\\

~~~~~~~~~~~~~~~~~~~~~~~~~~~~~~~~~~~~~~~~~~~~~~~~~~~~~~Fig.5~~~~~~~~~~~~~~~~~~~~\\
\vspace{1mm} \textit{\textbf{Fig.5} shows the effect of the
rainbow functions on the outcome of collapse for the Starobinsky's
model.}
\end{figure}

\subsection{Power Law Model}

\subsubsection{Case-1 $(n\geq 4)$}
Using equations (\ref{massfunc1}) and (\ref{X0}), we have
\begin{eqnarray}\label{X02}
\frac{2}{X_{0}}=
\begin{array}llim\\
\begin{tiny}t\rightarrow 0\end{tiny}\\
\begin{tiny}r\rightarrow 0\end{tiny}
\end{array}
\frac{\mathcal{G}(E)}{\mathcal{F}(E)}\left[1-\frac{f_{5}(t)}{r}+\frac{f_{6}(t)}{r^2}\right]
\end{eqnarray}
Just like the previous section we choose the following: \\\\
~~~$f_{5}(t)=\gamma_{2} t$,~~~~$f_{6}(t)=\delta_{2} t^{2}$\\\\
where $\gamma_{2}$ and $\delta_{2}$ are arbitrary constants. Using
the above functions in eqn.(\ref{X02}) we get the following
algebraic equation,
\begin{equation}\label{algebra2}
\delta_{2}X_{0}^{3}-\gamma_{2}X_{0}^{2}+X_{0}-\frac{2\mathcal{F}(E)}{\mathcal{G}(E)}=0
\end{equation}
Solving the above equation we get only one real root for $X_{0}$.
The other two roots are complex conjugates which are of no
interest in this study. The real root is given by,
\begin{equation}\label{root1}
X_{0}=\frac{1}{3\delta_{2}}\left[\gamma_{2}-\frac{2^{1/3}\left(3\delta_{2}-\gamma_{2}^{2}\right)}{P}+\frac{P}{2^{1/3}}\right],~~~~~~\delta_{2}\neq
0, ~~P\neq 0
\end{equation}
where
\begin{equation}
P=\left[2\gamma_{2}^{3}-9\gamma_{2}\delta_{2}
+\frac{54\mathcal{F}(E)\delta_{2}^{2}}{\mathcal{G}(E)}+\sqrt{4\left(3\delta_{2}-\gamma_{2}\right)^{3}+\left(2\gamma_{2}^{3}-9\gamma_{2}\delta_{2}
+\frac{54\mathcal{F}(E)\delta_{2}^{2}}{\mathcal{G}(E)}\right)^{2}}\right]^{1/3}
\end{equation}
Here $P$ may be positive or negative satisfying
$4\left(3\delta_{2}-\gamma_{2}\right)^{3}+\left(2\gamma_{2}^{3}-9\gamma_{2}\delta_{2}
+\frac{54\mathcal{F}(E)\delta_{2}^{2}}{\mathcal{G}(E)}\right)^{2}\geq
0$. \\
So the condition for a naked singularity becomes
\begin{equation}\label{nakedcond}
\gamma_{2}-\frac{2^{1/3}\left(3\delta_{2}-\gamma_{2}^{2}\right)}{P}+\frac{P}{2^{1/3}}>0,
~~~for ~~\delta_{2}>0
\end{equation}
and that for a black hole is,
\begin{equation}\label{bhcond}
\gamma_{2}-\frac{2^{1/3}\left(3\delta_{2}-\gamma_{2}^{2}\right)}{P}+\frac{P}{2^{1/3}}<0,
~~~for ~~\delta_{2}<0
\end{equation}
In the limit $\mathcal{F}(E)\rightarrow 1,
\mathcal{G}(E)\rightarrow 1$ we can retrieve the corresponding
conditions in general relativity.

\subsubsection{Case-2 $(n=2)$}
Using equations (\ref{massfunc1}) and (\ref{X0}), we have
\begin{eqnarray}\label{X03}
\frac{2}{X_{0}}=
\begin{array}llim\\
\begin{tiny}t\rightarrow 0\end{tiny}\\
\begin{tiny}r\rightarrow 0\end{tiny}
\end{array}
\frac{\mathcal{G}(E)}{\mathcal{F}(E)}\left[1-A\left(\frac{f_{7}(t)}{r}-\frac{f_{8}(t)}{r^2}\right)-B\left(\frac{f_{9}(t)}{r}-\frac{f_{10}(t)}{r^2}+f_{11}(t)r^{2}
+f_{12}(t)r^{3}\right)\right]
\end{eqnarray}
We choose the following: \\\\
$f_{7}(t)=\gamma_{3} t$, ~~~$f_{8}(t)=\delta_{3} t^{2}$,
~~~$f_{9}(t)=\gamma_{4} t$, ~~~$f_{10}(t)=\delta_{4} t^{2}$,
~~~$f_{11}(t)=\xi_{3} t^{-2}$, ~~~$f_{12}(t)=\epsilon_{3} t^{-3}$
\\\\
where $\gamma_{3}$, $\delta_{3}$, $\gamma_{4}$, $\delta_{4}$,
$\xi_{3}$ and $\epsilon_{3}$ are arbitrary constants. Using the
above functions in eqn.(\ref{X03}) we get the following algebraic
equation,
\begin{equation}\label{algebra3}
\tau X_{0}^{5}-\zeta
X_{0}^{4}+X_{0}^{3}-\frac{2\mathcal{F}(E)}{\mathcal{G}(E)}X_{0}^{2}-B\xi_{3}X_{0}-B\epsilon_{3}=0
\end{equation}
where $\tau=A\delta_{3}+B\delta_{4}$ and
$\zeta=A\gamma_{3}+B\gamma_{4}$. Since it is highly unlikely to
get a feasible solution of the above fifth degree equation, we
generate plots of $X_{0}$ against various parameters to get an
idea about the nature of $X_{0}$. The plots have been generated in
Figs. 6, 7, 8 and 9.

\begin{figure}
~~~~~~~~~~~\includegraphics[height=1.7in,width=2.2in]{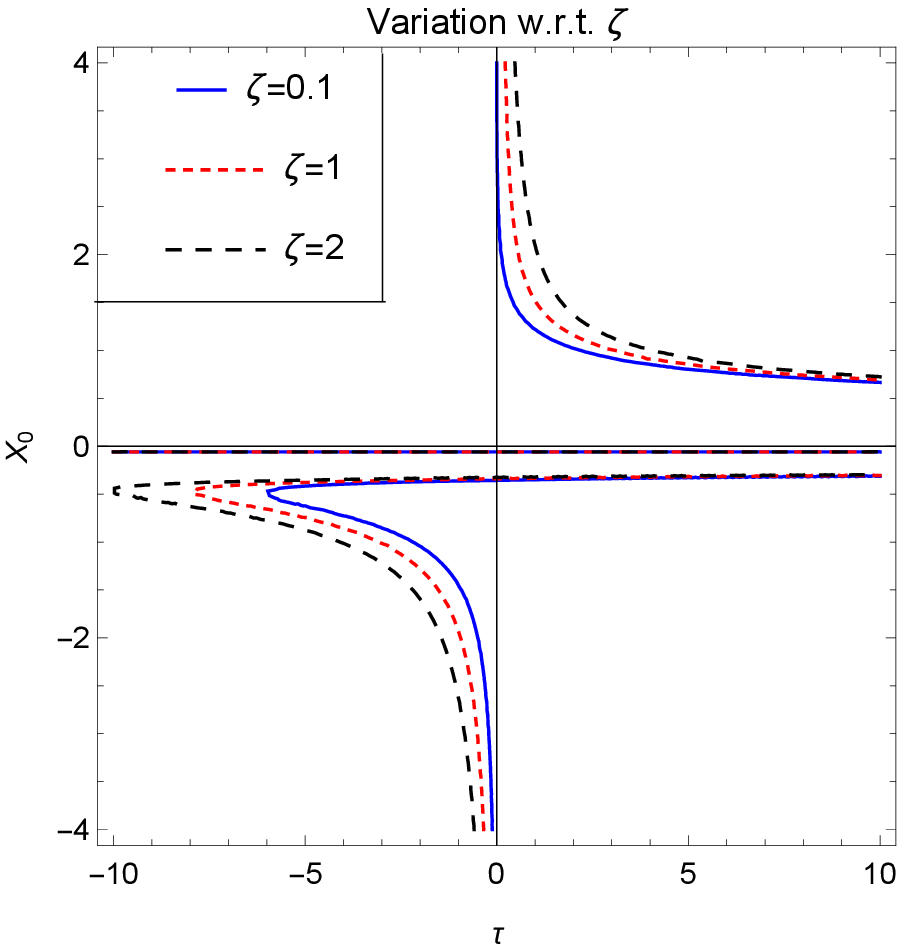}~~~~~~~~~~~~~\includegraphics[height=1.7in,width=2.2in]{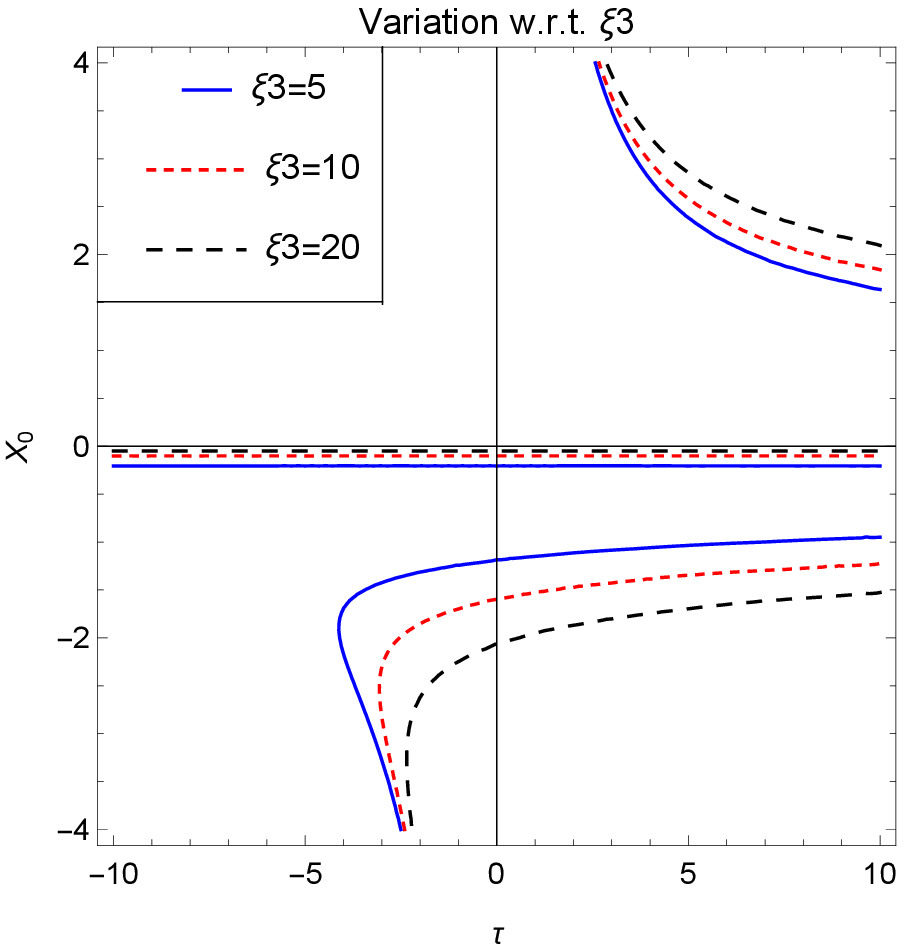}~~~~~~~\\

~~~~~~~~~~~~~~~~~~~~~~~~~~~~~~~Fig.6~~~~~~~~~~~~~~~~~~~~~~~~~~~~~~~~~~~~~~~~~~~~~~~~~~~~~~~~~Fig.7~~~~~~~~~\\

~~~~~~~~~~~\includegraphics[height=1.7in,width=2.2in]{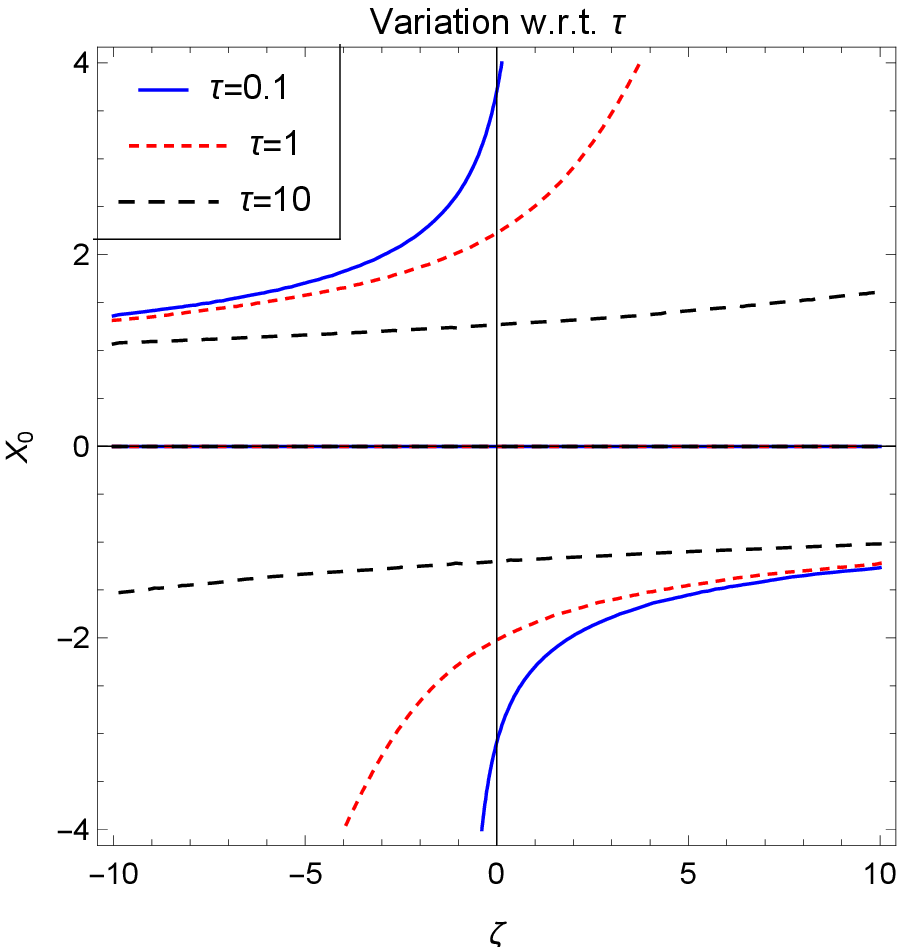}~~~~~~~~~~~~~\includegraphics[height=1.7in,width=2.2in]{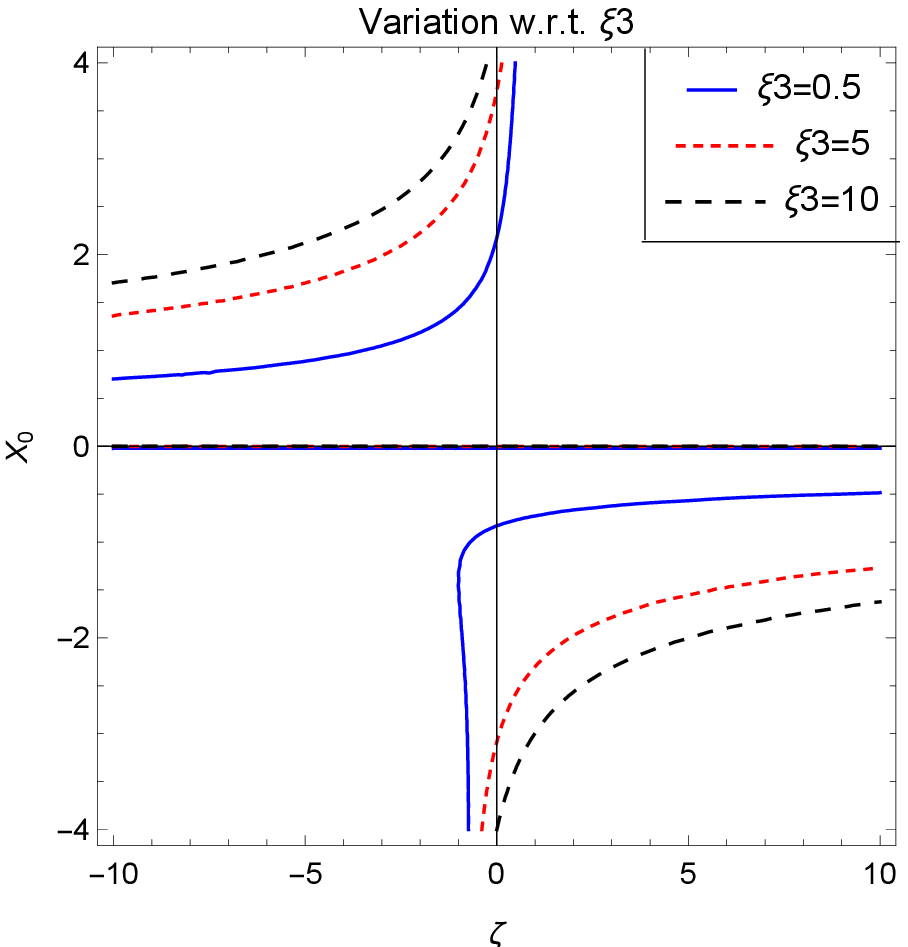}~~~~~~~\\

~~~~~~~~~~~~~~~~~~~~~~~~~~~~Fig.8~~~~~~~~~~~~~~~~~~~~~~~~~~~~~~~~~~~~~~~~~~~~~~~~~~~~~~~~~~~~Fig.9~~~~~~~~~\\

\vspace{1mm} \textit{\textbf{Figs.6 and 7} show the variation of
$X_{0}$ with $\tau$ for different values of $\zeta$ and $\xi_3$
respectively for the Power law model (Case-2). In Fig.6 the
initial conditions are taken as $\xi_3=20$, $\epsilon_3=1$,
$B=0.05$, $\eta=1$, $n=2$, $E_{s}=1$ and $E_{P}=5$. In Fig.7 the
initial conditions are $\zeta=10$, $\epsilon_3=1$, $B=5$,
$\eta=1$, $n=2$, $E_{s}=1$ and $E_{P}=5$.}\\\\
\vspace{1mm} \textit{\textbf{Figs.8 and 9} show the variation of
$X_{0}$ with $\zeta$ for different values of $\tau$ and $\xi_3$
respectively for the Power law model (Case-2). In Fig.8 the
initial conditions are taken as $\xi_3=5$, $\epsilon_3=0.01$,
$B=5$, $\eta=1$, $n=2$, $E_{s}=1$ and $E_{P}=5$. In Fig.9 the
initial conditions are $\tau=0.1$, $\epsilon_3=0.01$, $B=5$,
$\eta=1$, $n=2$, $E_{s}=1$ and $E_{P}=5$.}
\end{figure}

\begin{figure}
~~~~~~~~~~~~~~~~~~~~~~~~~~\includegraphics[height=2.5in,width=2.5in]{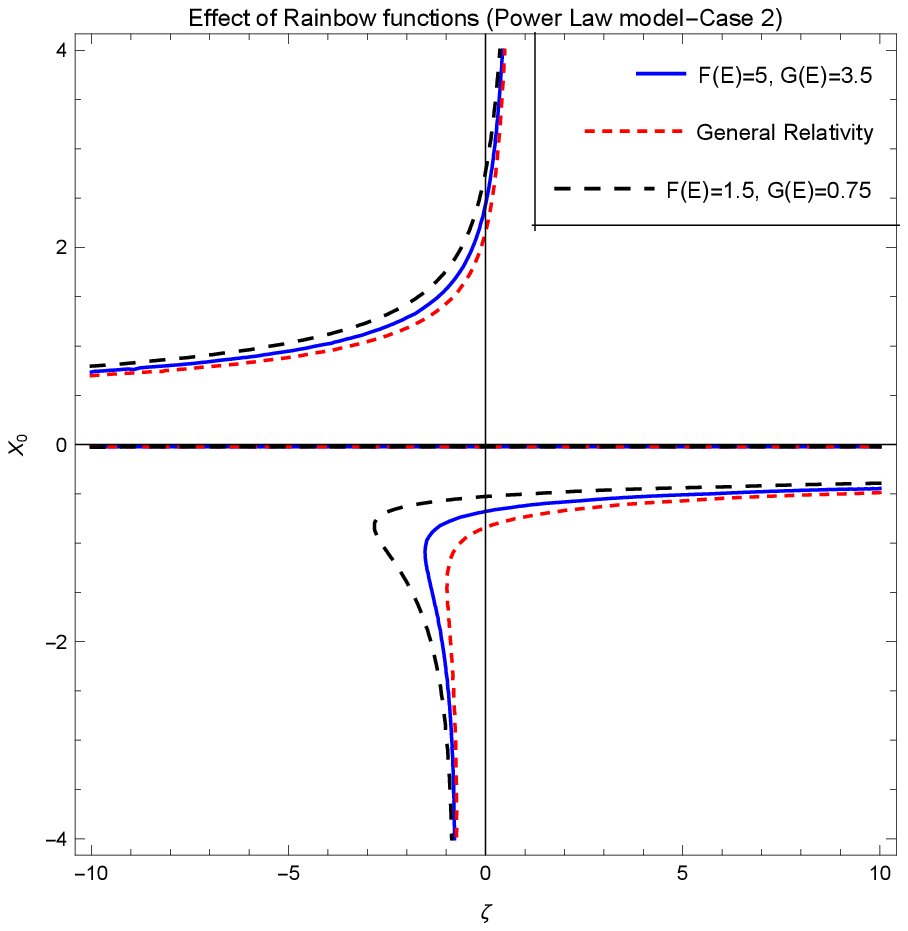}~~~~~~~~\\

~~~~~~~~~~~~~~~~~~~~~~~~~~~~~~~~~~~~~~~~~~~~~~~~~~~~~~Fig.10~~~~~~~~~~~~~~~~~~\\

\vspace{1mm} \textit{\textbf{Fig.10} shows the effect of the
rainbow functions on the outcome of collapse for the power law
model (case-2).}
\end{figure}

\subsection{Numerical Analysis of the Results} Here we will analyze the numerical
results obtained in the previous section. In case of the
Starobinsky's model we obtained a five degree equation in $X_{0}$
given by eqn.(\ref{algebra1}). Unable to get a general solution by
the known algebraic methods, we have resorted to numerical
techniques to get an idea of the nature of $X_{0}$, which in turn
determines the nature of singularity formed. In Figs.1 and 2 we
have generated contour plots for $X_{0}$ against $\delta_{1}$ for
different values of $\gamma_{1}$ and $\xi_1$ respectively. From
Fig.1 we see that for $\delta_{1}<0$, $X_{0}$ remains in the
negative level for different values of $\gamma_{1}$, thus
favouring the formation of a BH. But for $\delta_{1}>0$, we see
trajectories of $X_{0}$ both in the negative and positive levels.
This indicates that there is a realistic chance of formation of NS
for this range. In Fig.2, we have almost a similar result as in
Fig.1. In the both the figures we see that the tendency of
formation of NS increases with the increase in the value of the
parameter. So the dependencies of collapsing procedure on
$\gamma_{1}$ and $\xi_1$ are almost of the identical nature.

In Figs.3 and 4 plots are generated for $X_{0}$ against $\xi_1$
for different values of $\delta_{1}$ and $\gamma_{1}$
respectively. In Fig.3 we see that almost all the trajectories lie
in the positive region throughout the entire domain of $\xi_1$,
thus favouring the formation of NS. Only around $\xi_1\geq 8$, we
see that there is a transition and the possibility of the
formation of an event horizon brightens. It should also be
mentioned here that an increase in the value of $\delta_{1}$
decreases the chance of formation of NS. A near similar trend is
obtained in Fig.4 where we see that the tendency of formation of
NS increases with the increase in $\gamma_{1}$. Moreover the
transition from NS to BH starts at an earlier stage around
$\xi_1=3$ as compared to Fig.3. The study would lose its
significance if we cannot understand the role played by gravity's
rainbow in the collapsing scenario. This is shown in Fig.5, where
we clearly see the trajectories for different rainbow functions
and can also compare with that of general relativity. It can be
seen that gravity's rainbow has a tendency to push the
trajectories towards the positive level thus indicating greater
affinity towards the formation of NS.

In the case-2 $(n=2)$ of the power law model we encounter another
five degree algebraic equation in $X_{0}$ given in
eqn.(\ref{algebra3}). Figs.6, 7, 8 and 9 are dedicated to the
study of this equation. In Figs.6 and 7, we have generated plots
for $X_{0}$ against $\tau$ for different values of $\zeta$ and
$\xi_3$ respectively. In Fig.6, for $\tau<0$, any singularity that
forms will be clothed by an event horizon, thus forming a BH. On
the contrary, for $\tau>0$ we see that there is a fair chance of
formation of NS. Eventually what type of singularity forms depends
on the initial conditions for $\tau>0$. Similar results are
visible in Fig.7, where trajectories are obtained for different
values of $\xi_3$. In both the plots the tendency of NS increases
with the increase in the values of parameters $\zeta$ and $\xi_3$.

In Figs.8 and 9, plots have been obtained for $X_{0}$ against the
parameter $\zeta$ for different values of $\tau$ and $\xi_3$
respectively. In Fig.8, we see that throughout the domain of
$\zeta$ trajectories exist both in the positive and negative
level. So tendency of formation of both BH and NS exists for any
value of $\zeta$ and eventually depends on the initial conditions
for the final outcome. But with the increase in the value of
$\tau$ the tendency of formation of NS decreases. From Fig.9 it is
evident that for $\xi_3<0$, collapse results in the formation of
NS, whereas for $\xi_3>0$, the generic tendency is the formation
of BH. In Fig.10, we can see the effect of gravity's rainbow on
the collapsing process, which is same as the previous case.
Gravity's rainbow increases the tendency of formation of NS.

In almost all the figures we see that only one trajectory appears
at the positive level, thus indicating that only one null geodesic
escapes the singularity thus making it locally naked in nature. In
Fig.4 we find a striking difference from the above fact. We see
that in the $\xi_{1}<0$ region we have three positive roots of
$X_{0}$ for $\gamma_{1}=3$. This would mean that the singularity
formed in this case is visible to a distant observer for a
sufficiently long amount of time to make it globally naked in
nature. So the Starobinsky's model supports a globally naked
singularity. Understanding the role of initial data, it must be
mentioned that any asymptotic nature near the vertical axis
(corresponds to a bundle of null geodesics escaping from the
singularity) would eventually result in global nakedness which may
be the case in Figs.1, 2 and 6.

\subsection{Strength of the singularity (Curvature growth near the
singularity)} The strength of singularity is defined as the
measure of its destructive capacity. The prime concern is that
whether extension of space-time is possible through the
singularity or not under any situation. Following Tipler
\cite{Tipler} a curvature singularity is said to be strong if any
object hitting it is crushed to zero volume. In \cite{Tipler} the
condition for a strong singularity is given by,
\begin{eqnarray}
\begin{array}{c}
S=\lim \tau^{2}\psi \\
\begin{tiny}\tau\rightarrow 0\end{tiny}\\
\end{array}
\begin{array}{c}
=\lim \tau^{2}R_{\mu\nu}K^{\mu}K^{\nu}>0 \\
\begin{tiny}\tau\rightarrow 0\end{tiny}\\
\end{array}
\end{eqnarray}
where $R_{\mu\nu}$ is the Ricci tensor, $\psi$ is a scalar given
by $\psi=R_{\mu\nu}K^{\mu}K^{\nu}$, where
$K^{\mu}=\frac{dx^{\mu}}{d\tau}$ is the tangent to the non
spacelike geodesics at the singularity and $\tau$ is the affine
parameter. In the paper \cite{Maharaj} Mkenyeleye et al. have
shown that,
\begin{eqnarray}\label{maha}
\begin{array}{c}
S=\lim \tau^{2}\psi \\
\begin{tiny}\tau\rightarrow 0\end{tiny}\\
\end{array}
\begin{array}{c}\label{stren}
=\frac{1}{4}X_{0}^{2}\left(2\dot{m_{0}}\right) \\
\begin{tiny}~\end{tiny}\\
\end{array}
\end{eqnarray}
where
\begin{eqnarray}
\begin{array}{c}
m_{0}=\lim~ m(t,r) \\
\begin{tiny}t\rightarrow 0\end{tiny}\\
\begin{tiny}r\rightarrow 0\end{tiny}
\end{array}
\end{eqnarray}
and
\begin{eqnarray}\label{massd}
\begin{array}{c}
\dot{m_{0}}=\lim \frac{\partial}{\partial~t}\left(m(t,r)\right) \\
\begin{tiny}t\rightarrow 0\end{tiny}\\
\begin{tiny}r\rightarrow 0\end{tiny}
\end{array}
\end{eqnarray}
In ref. \cite{Maharaj} it has also been shown that the relation
between $X_{0}$ and the limiting values of mass is given by,
\begin{equation}\label{xmass}
X_{0}=\frac{2}{1-2m_{0}'-2\dot{m_{0}}X_{0}}
\end{equation}
where
\begin{eqnarray}\label{dashedmass}
\begin{array}{c}
m_{0}'=\lim \frac{\partial}{\partial~r}\left(m(t,r)\right) \\
\begin{tiny}t\rightarrow 0\end{tiny}\\
\begin{tiny}r\rightarrow 0\end{tiny}
\end{array}
\end{eqnarray}
and $\dot{m_{0}}$ is given by the eqn.(\ref{massd}). It was shown
by Dwivedi and Joshi in Ref.\cite{strongcurvature, cch2} that a
classical singularity in Vaidya spacetime is supposed to be a
strong curvature singularity in a very strong sense. It was also
shown that the conjecture \cite{tiplernew} that the strong
curvature singularities are never naked is not true. Moreover the
structure of such NS were studied in detail in
Ref.\cite{strongcurvature1} and it was shown that the singularity
presents a directional behaviour in terms of curvature growth
along the singular geodesics. In a quantum regime the singularity
formed is supposed to become gravitationally weak, thus allowing a
continuous extension of the spacetime beyond the singularity
\cite{weakness}. Below we study the strength of the singularities
for the different models.

\subsubsection{Starobinsky's Model} Using eqn.(\ref{mass2}) in the
above relation (\ref{maha}) we get
\begin{eqnarray}\label{strength2}
\begin{array}{c}
S=\lim \tau^{2}\psi \\
\begin{tiny}\tau\rightarrow 0\end{tiny}\\
\end{array}
\begin{array}{c}
=\frac{1}{2}X_{0}^{2}\left[\gamma_{1}-2\delta_{1}X_{0}-\frac{2\xi_{1}}{X_{0}^{3}}-\frac{3\epsilon_{1}}{X_{0}^{4}}\right] \\
\begin{tiny}~\end{tiny}\\
\end{array}
\end{eqnarray}
Using eqns.(\ref{mass2}), (\ref{massd}) and (\ref{dashedmass}) in
eqn.(\ref{xmass}) we get an equation for $X_{0}$
\begin{equation}\label{s1}
2\delta_{1}X_{0}^{5}-2\gamma_{1}X_{0}^{4}+X_{0}^{3}-2X_{0}^{2}-2\xi_{1}
X_{0}-2\epsilon_{1}=0
\end{equation}
The above equation can be investigated for roots and using these
values of $X_{0}$ in the eqn.(\ref{strength2}) we get the
conditions for which $S=\lim \tau^{2}\psi >0$, i.e. the conditions
under which we get a strong singularity. For a strong singularity
we have from eqn.(\ref{strength2})
\begin{equation}\label{cond1}
2\delta_{1}X_{0}^{5}-\gamma_{1}X_{0}^{4}+2\xi_{1}X_{0}+3\epsilon_{1}<0
\end{equation}
where $X_{0}$ can be obtained as solutions of eqn.(\ref{s1}).
Since eqn.(\ref{s1}) is a fifth degree algebraic equation in
$X_{0}$, it is difficult to get a general solution for $X_{0}$ by
the known mathematical methods. However we can get various
solutions numerically. Below we give a particular example of
this.\\

\textbf{Example}\\\\
Here we find out the roots of eqn.(\ref{s1}) for a particular set
of values of the parameters involved. Considering $\gamma_{1}=1$,
$\delta_{1}=0.1$, $\xi_{1}=0.5$, $\epsilon_{1}=10$, the roots of
the eqn.(\ref{s1}) are found to be $X_{0}=-1.0636-1.23735~i,~~
-1.0636+1.23735~i,~~ 1.26098-1.52332~i,~~ 1.26098+1.52332~i,~~
9.60523$. Taking the real value of $X_{0}=9.60523$ and the above
values of the parameters we get $S=-42.7027<0$ from
eqn.(\ref{strength2}). Hence for this particular set of values the
singularity is weak. Similarly for a different set of initial
conditions we may have different results.

\subsubsection{Power Law Model}

\textbf{Case-1 $(n\geq 4)$} \\\\
Using eqn.(\ref{massfunc1}) in the above relation (\ref{maha}) we
get,
\begin{eqnarray}\label{strength3}
\begin{array}{c}
S=\lim \tau^{2}\psi \\
\begin{tiny}\tau\rightarrow 0\end{tiny}\\
\end{array}
\begin{array}{c}
=\frac{1}{2}X_{0}^{2}\left[\gamma_{2}-2\delta_{2}X_{0}\right] \\
\begin{tiny}~\end{tiny}\\
\end{array}
\end{eqnarray}
Using eqns.(\ref{massfunc1}), (\ref{massd}) and (\ref{dashedmass})
in eqn.(\ref{xmass}) we get an equation for $X_{0}$,
\begin{equation}\label{s2}
2\delta_{2}X_{0}^{3}-2\gamma_{2}X_{0}^{2}+X_{0}-2=0
\end{equation}
Solving the above equation we get only one real root given by,
\begin{equation}\label{rootpowerlaw}
X_{0}=\frac{\gamma_{2}}{3\delta_{2}}-\frac{6\delta_{2}-4\gamma_{2}^{2}}{6\times2^{1/3}\delta_{2}Q^{1/3}}+\frac{Q^{1/3}}{3\times2^{2/3}\delta_{2}}
\end{equation}
where~~
$Q=4\gamma_{2}^{3}-9\gamma_{2}\delta_{2}+54\delta_{2}^{2}+3\sqrt{3\left(108\delta_{2}^{4}-36\gamma_{2}\delta_{2}^{3}+2\delta_{2}^{3}+16\gamma_{2}^{3}\delta_{2}^{2}
-\gamma_{2}^{2}\delta_{2}^{2}\right)}$.\\\\
Using the value of $X_{0}$ from eqn.(\ref{rootpowerlaw}) in
eqn.(\ref{strength3}) we get the condition for a strong
singularity as,
\begin{equation}\label{strongcond}
\frac{\left(-2\times2^{2/3}\gamma_{2}^{2}+3\times2^{2/3}\delta_{2}+\gamma_{2}Q^{1/3}-2^{1/3}Q^{2/3}\right)\left(2\times2^{2/3}\gamma_{2}^{2}-3\times2^{2/3}\delta_{2}
+2\gamma_{2}Q^{1/3}+2^{1/3}Q^{1/3}\right)^{2}}{216\delta_{2}^{2}Q}>0
\end{equation}

\textbf{Case-2 $(n=2)$}\\\\
Using eqn.(\ref{massfunc1}) in the above relation (\ref{maha}) we
get,
\begin{eqnarray}\label{strength4}
\begin{array}{c}
S=\lim \tau^{2}\psi \\
\begin{tiny}\tau\rightarrow 0\end{tiny}\\
\end{array}
\begin{array}{c}
=\frac{1}{2}X_{0}^{2}\left[A\left(\gamma_{3}-2\delta_{3}X_{0}\right)+B\left(\gamma_{4}-2\delta_{4}X_{0}-\frac{2\xi_{3}}{X_{0}^{3}}-\frac{3\epsilon_{3}}{X_{0}^{4}}\right)\right] \\
\begin{tiny}~\end{tiny}\\
\end{array}
\end{eqnarray}
Using eqns.(\ref{massfunc1}), (\ref{massd}) and (\ref{dashedmass})
in eqn.(\ref{xmass}) we get an equation for $X_{0}$,
\begin{equation}\label{s3}
2\left(A\delta_{3}+B\delta_{4}\right)X_{0}^{5}-2\left(A\gamma_{3}+B\gamma_{4}\right)X_{0}^{4}+X_{0}^{3}-2X_{0}^{2}-2B\xi_{3}X_{0}-2B\epsilon_{3}=0
\end{equation}
Just like the Starobinsky's model, it is difficult to obtain a
general solution for $X_{0}$ from the above equation. So we will
resort to a numerical solution and give a particular example of
it.\\\\
\textbf{Example}\\\\
Here we consider the following set of numerical values for the
parameters involved,\\
$\gamma_{3}=1,~~\gamma_{4}=0.1,~~\delta_{3}=0.1,~~\delta_{4}=0.5,~~\xi_{3}=0.5,~~\epsilon_{3}=10,~~A=2,~~B=5$\\
Using these values in eqn.(\ref{s3}) we get the roots as,\\
$X_{0}=-1.26336-1.03348~i,~~-1.26336+1.03348~i,~~0.6967-1.70004~i,~~0.6967+1.70004~i,~~2.05924$.
Taking the real value of $X_{0}=2.05924$ and the above values of
the parameters we get $S=-37.177<0$ from eqn.(\ref{strength4}).
Hence for this particular set of values the singularity is weak.
But by changing the initial conditions we may have different
results.

Using eqn.(\ref{strength4}) we get the condition for a strong
singularity for this model as,
\begin{equation}\label{strongcond2}
A\left(\gamma_{3}-2\delta_{3}X_{0}\right)+B\left(\gamma_{4}-2\delta_{4}X_{0}-\frac{2\xi_{3}}{X_{0}^{3}}-\frac{3\epsilon_{3}}{X_{0}^{4}}\right)>0
\end{equation}
where $X_{0}$ is obtained from eqn.(\ref{s3}).

\section{A Thermodynamical Analysis of the System}
Here we will study the thermodynamical properties of the models.
Thermodynamics is the heart of any physical process which deals
with exchange of heat to and from the system. So this study is
crucial in our analysis as it involves a collapsing mechanism and
eventually the formation of a singularity which can be either a BH
or NS. It is a well known fact that BH thermodynamics is a very
important topic of astrophysics. In a fundamental theory of
quantum gravity, it is expected that the thermodynamic properties
of a system should emerge from a microscopic statistical
description. Deriving motivation from these facts we proceed to
study the thermodynamical aspects of the BHs formed as a result of
collapse for the various models.

\subsection{Starobinsky's Model}
The event horizon ($r_{h}$) can be obtained from the relation
$f(t,r)=0$, i.e.,
\begin{equation}\label{eventhorizon}
-\frac{1}{\mathcal{F}^2(E)}\left[1-\frac{1}{r}\left(f_{1}(t)-\frac{f_{2}(t)}{r}+f_{3}(t)r^{3}+f_{4}(t)r^{4}\right)\right]=0
\end{equation}
The real positive root of the above equation gives the radius of
the event horizon. The thermalization temperature of a system is
defined as the temperature at which the system attains thermal
equilibrium. The relation for thermalization temperature is given
by,
\begin{equation}\label{thermaltemp}
T=\frac{1}{4\pi}\frac{d}{dr}f(t,r)|_{r=r_{h}}
\end{equation}
For the Starobinsky's model the expression for thermalization
temperature becomes,
\begin{equation}\label{thermaltemp1}
T=\frac{1}{4\mathcal{F}^{2}(E)\pi
r^{3}}\left[-rf_{1}(t)+2f_{2}(t)+r^{4}\left(2f_{3}(t)+3rf_{4}(t)\right)\right]
\end{equation}
The entropy of the system is given by,
\begin{equation}\label{entropy}
S=\pi^{2}r_{h}^{2}
\end{equation}
where we consider $\pi G=1$. Here $G$ being the Newton's universal
gravitation constant. Total energy can be obtained from the
relation
\begin{equation}\label{totenergy}
U=\int T~ dS
\end{equation}
Using the equations (\ref{thermaltemp1}), (\ref{entropy}) and
(\ref{totenergy}) we get the total energy for this model as,
\begin{equation}\label{totenergy1}
U=\frac{\pi}{24\mathcal{F}^{2}(E)r}\left[-12rf_{1}(t)\log(r)-24f_{2}(t)+8r^{4}f_{3}(t)+9r^{5}f_{4}(t)\right]
\end{equation}
Helmholtz free energy is a thermodynamic potential which is the
measure of the useful work obtainable from a closed system at
constant temperature and volume. It is given by the relation,
\begin{equation}\label{helmholtz}
F_{1}=U-TS
\end{equation}
Using the eqns.(\ref{thermaltemp1}), (\ref{entropy}) and
(\ref{totenergy1}) we get the expression Helmholtz free energy for
this model as,
\begin{equation}\label{helmholtz1}
F_{1}=-\frac{\pi}{24\mathcal{F}^{2}(E)r}\left[6rf_{1}(t)\left\{2\log(r)-1\right\}+36f_{2}(t)+r^{4}\left\{4f_{3}(t)+9rf_{4}(t)\right\}\right]
\end{equation}
Specific heat at constant volume is given by,
\begin{equation}\label{specificheat}
C=\left(\frac{dU}{dT}\right)_{V}
\end{equation}
Using relations (\ref{thermaltemp1}), (\ref{entropy}),
(\ref{totenergy}) and (\ref{specificheat}) we get the specific
heat at constant volume for the Starobinsky's model as,
\begin{equation}\label{specificheat1}
C=\frac{\pi^{2}r^{2}\left[-rf_{1}(t)+2f_{2}(t)+r^{4}\left\{2f_{3}(t)+3rf_{4}(t)\right\}\right]}{rf_{1}(t)-3f_{2}(t)+r^{4}\left\{f_{3}(t)+3rf_{4}(t)\right\}}
\end{equation}
In Figs.11, 12, 13 and 14 the thermodynamical parameters have been
plotted against the radial coordinate $r$ for the Starobinsky's
model. In this section we have used the rainbow functions given by
eqn.(\ref{rainbowfuncchoice1}). In the thermodynamic parameters it
is seen that the rainbow function present is $\mathcal{F}(E)$. So
if we use the rainbow functions given by
eqn.(\ref{rainbowfuncchoice}) as used in the collapse study, the
results would correspond to that of GR (since $\mathcal{F}(E)=1$).
This is the reason why we choose a different set of values for the
rainbow functions for the thermodynamic study. Moreover by using
different values of $\mathcal{F}(E)$ we can understand the effect
of gravity's rainbow on the thermodynamics of the system and also
compare with the results of GR by taking $\mathcal{F}(E)=1$. From
Fig.11 it is seen that the range of thermalization temperature $T$
increases with the increase in the value of $\delta_1$. On the
contrary in Fig.12 it is seen that the range of Internal Energy
$U$ decreases with increase in the value of $\delta_{1}$. In
Fig.13 the trend for Helmholtz free energy $F_1$ is obtained
against $r$. From the figure it is evident that with the increase
in $\delta_1$, there is a decrease in the value of $F_1$. In
Fig.14 a plot for the specific heat at constant volume $C$ is
obtained against $r$. From the figure we see that the curves are
discontinuous and each of them have two branches. We see that $C$
increases with the increase in $\delta_1$. Finally Fig.15 shows
the effect of gravity's rainbow on the thermodynamic parameter
$T$.

\begin{figure}
~~~~~~~~~~\includegraphics[height=1.7in,width=2.2in]{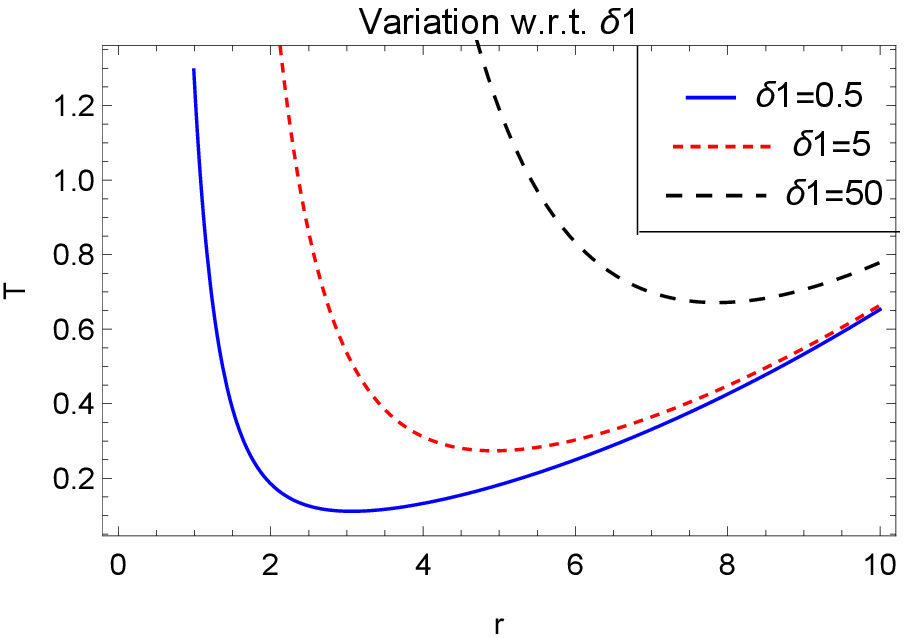}~~~~~~~~~~~~~\includegraphics[height=1.7in,width=2.2in]{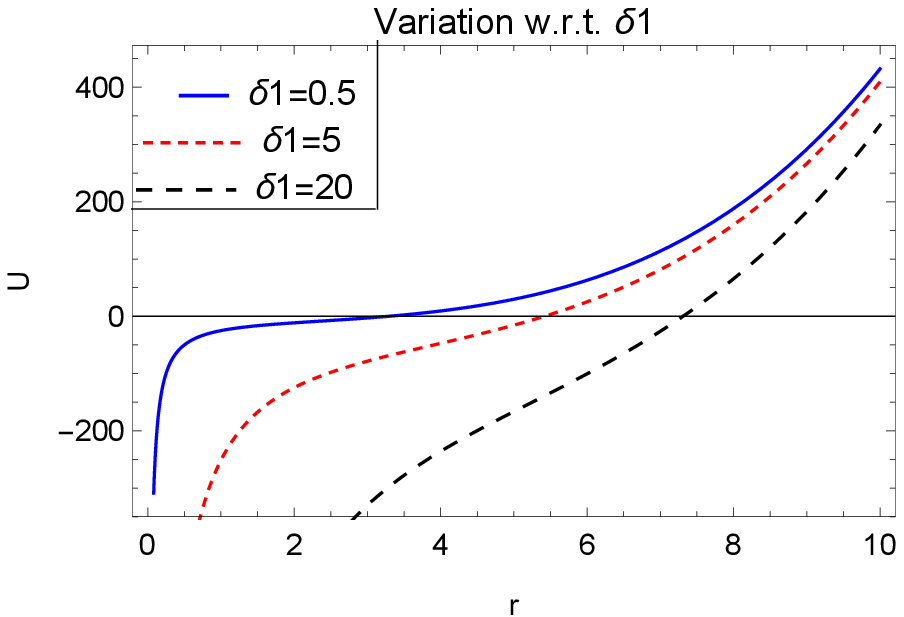}~~~~~~~\\

~~~~~~~~~~~~~~~~~~~~~~~~~~~~Fig.11~~~~~~~~~~~~~~~~~~~~~~~~~~~~~~~~~~~~~~~~~~~~~~~~~~~~~~~~~~~~Fig.12~~~~~~~~~\\

~~~~~~~~~~\includegraphics[height=1.7in,width=2.2in]{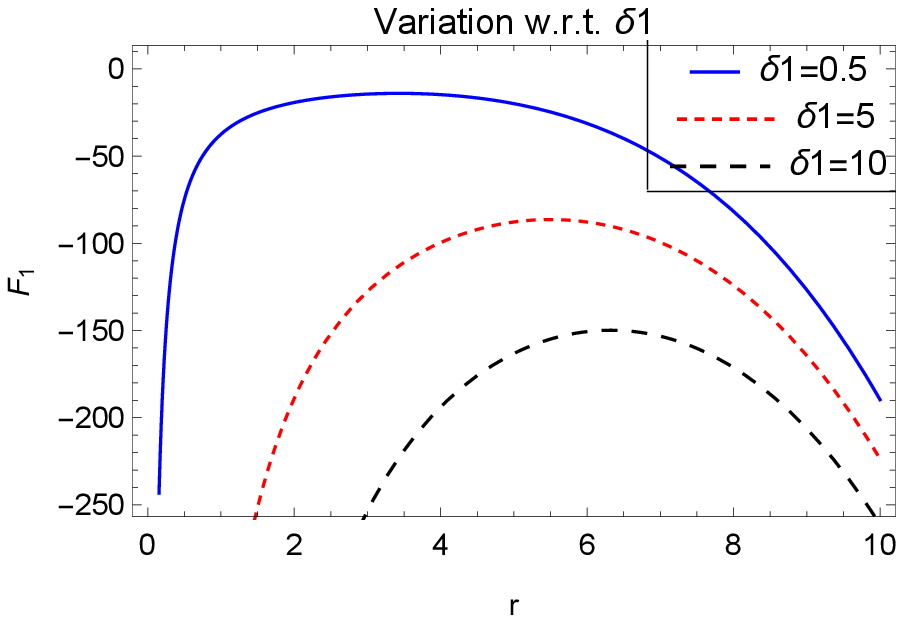}~~~~~~~~~~~~~\includegraphics[height=1.7in,width=2.2in]{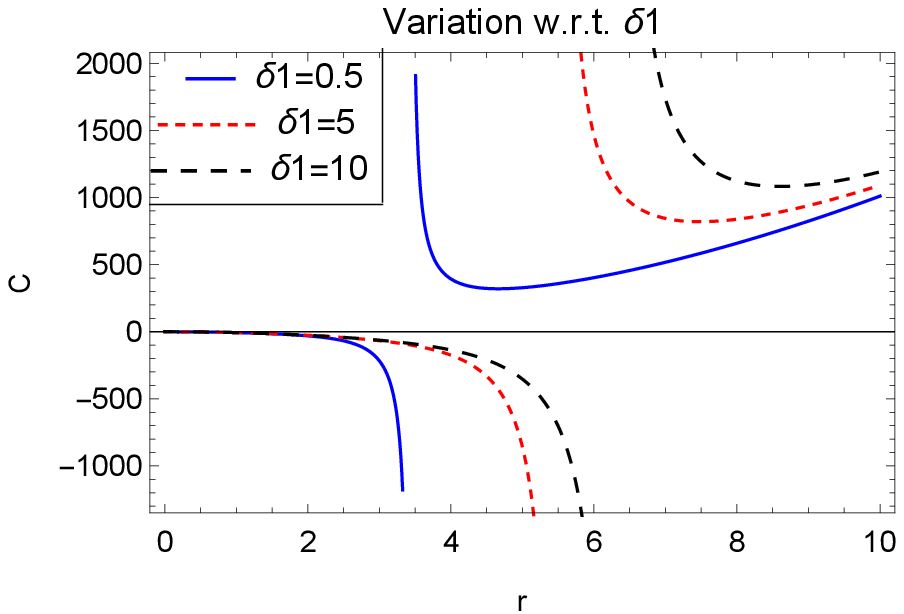}~~~~~~~\\

~~~~~~~~~~~~~~~~~~~~~~~~~~~~Fig.13~~~~~~~~~~~~~~~~~~~~~~~~~~~~~~~~~~~~~~~~~~~~~~~~~~~~~~~~~~~~Fig.14~~~~~~~~~\\

\vspace{1mm} \textit{\textbf{Figs.11, 12, 13 and 14} show the
variation of Thermalization temperature $(T)$, Internal Energy
$(U)$, Helmholtz free energy $(F_{1})$ and Specific Heat at
constant volume $(C)$ respectively with $r$ for different values
of $\delta_1$ for the Starobinsky's model. The rainbow function is
considered as $\mathcal{F}(E)=\frac{1}{1-a_{1}(E_{s}/E_{P})}$. In
Fig.11 the initial conditions are taken as $\gamma_1=0.1$,
$\xi_1=1$, $\epsilon_1=5$, $t=5$, $a_{1}=1$, $E_{s}=1$, $E_{P}=5$.
In Fig.12 the initial conditions are $\gamma_1=0.1$, $\xi_1=5$,
$\epsilon_1=5$, $t=5$, $a_{1}=1$, $E_{s}=1$, $E_{P}=5$. In both
the figures 13 and 14 the initial conditions are taken as
$\gamma_1=0.1$, $\xi_1=0.3$, $\epsilon_1=3$, $t=5$, $a_{1}=1$,
$E_{s}=1$, $E_{P}=5$.}
\end{figure}

\begin{figure}
~~~~~~~~~~~~~~~~~~~~~~~~~~\includegraphics[height=2in,width=2.5in]{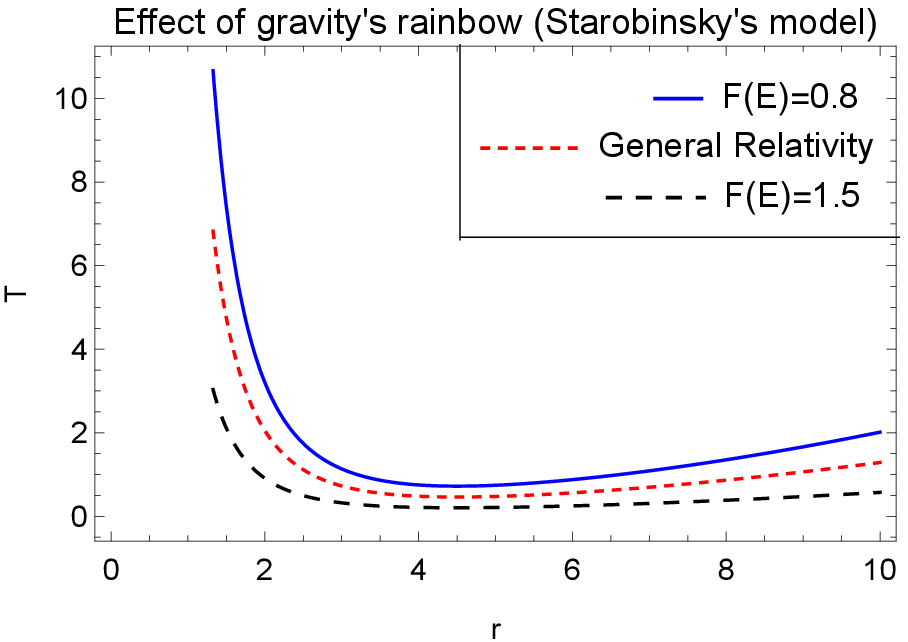}~~~~~~~~\\

~~~~~~~~~~~~~~~~~~~~~~~~~~~~~~~~~~~~~~~~~~~~~~~~~Fig.15~~~~~~~~~~~~~~~~~~\\

\vspace{1mm} \textit{\textbf{Fig.15} shows the effect of the
rainbow functions on Thermalization temperature for the
Starobinsky's model.}
\end{figure}

\subsection{Power Law Model}

\subsubsection{Case-1 $(n\geq 4)$}
The event horizon ($r_{h}$) can be obtained from the relation
$f(t,r)=0$, i.e.,
\begin{equation}\label{eventhorizon2}
-\frac{1}{\mathcal{F}^2(E)}\left[1-\frac{1}{r}\left(f_{5}(t)-\frac{f_{6}(t)}{r}\right)\right]=0
\end{equation}
The real positive root of the above equation gives the radius of
the event horizon. Solving the above equation we get,
\begin{equation}\label{radiushorizon}
r_{h}=\frac{1}{2}\left[f_{5}(t)\pm\sqrt{\left\{f_{5}(t)\right\}^{2}-4f_{6}(t)}\right]
\end{equation}
Obviously for a realistic event horizon radius we should have:\\\\
1. $\left\{f_{5}(t)\right\}^{2}\geq 4f_{6}(t)$ \\\\
2. $f_{5}(t)\pm\sqrt{\left\{f_{5}(t)\right\}^{2}-4f_{6}(t)}>0
\implies f_{6}>0$.\\

For this model the expression for thermalization temperature
becomes,
\begin{equation}\label{thermaltemp2}
T=\frac{1}{4\mathcal{F}^{2}(E)\pi
r^{3}}\left[2f_{6}(t)-rf_{5}(t)\right]
\end{equation}
Using the equations (\ref{thermaltemp2}), (\ref{entropy}) and
(\ref{totenergy}) we get the total energy for this model as,
\begin{equation}\label{totenergy2}
U=-\frac{\pi}{2\mathcal{F}^{2}(E)}\left[\frac{2f_{6}(t)}{r}+f_{5}(t)\log(r)\right]
\end{equation}
Using the eqns.(\ref{thermaltemp2}), (\ref{entropy}) and
(\ref{totenergy2}) we get the expression Helmholtz free energy for
this model as,
\begin{equation}\label{helmholtz2}
F_{1}=\frac{\pi}{4\mathcal{F}^{2}(E)r}\left[f_{5}(t)\left\{r-2r\log(r)\right\}-6f_{6}(t)\right]
\end{equation}
Finally using relations (\ref{thermaltemp2}), (\ref{entropy}),
(\ref{totenergy2}) and (\ref{specificheat}) we get the specific
heat at constant volume for this model as,
\begin{equation}\label{specificheat2}
C=\frac{\pi^{2}r^{2}\left[-rf_{5}(t)+2f_{6}(t)\right]}{rf_{5}(t)-3f_{6}(t)}
\end{equation}

\begin{figure}
~~~~~~~~~~\includegraphics[height=1.7in,width=2.2in]{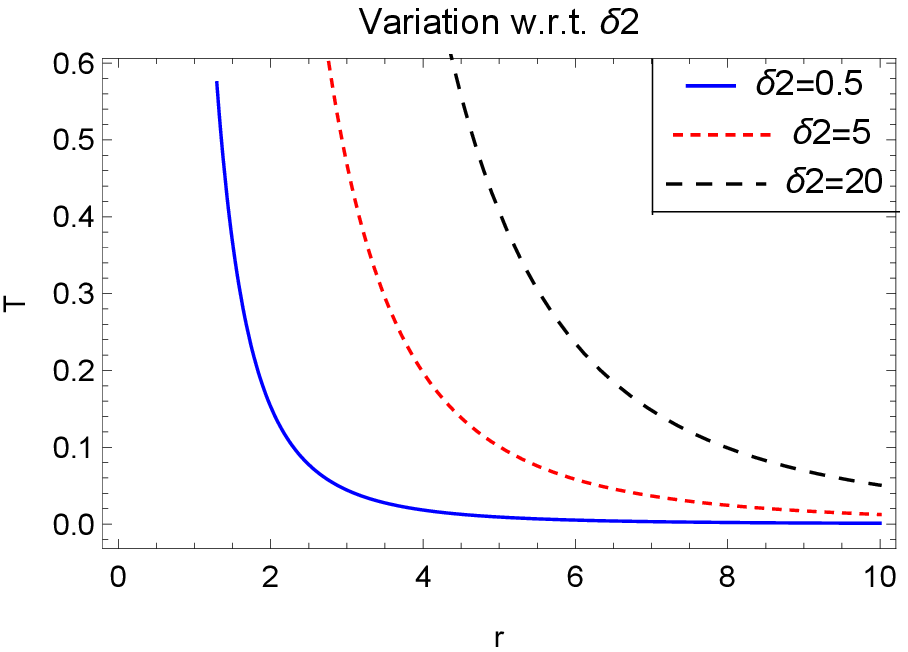}~~~~~~~~~~~~~\includegraphics[height=1.7in,width=2.2in]{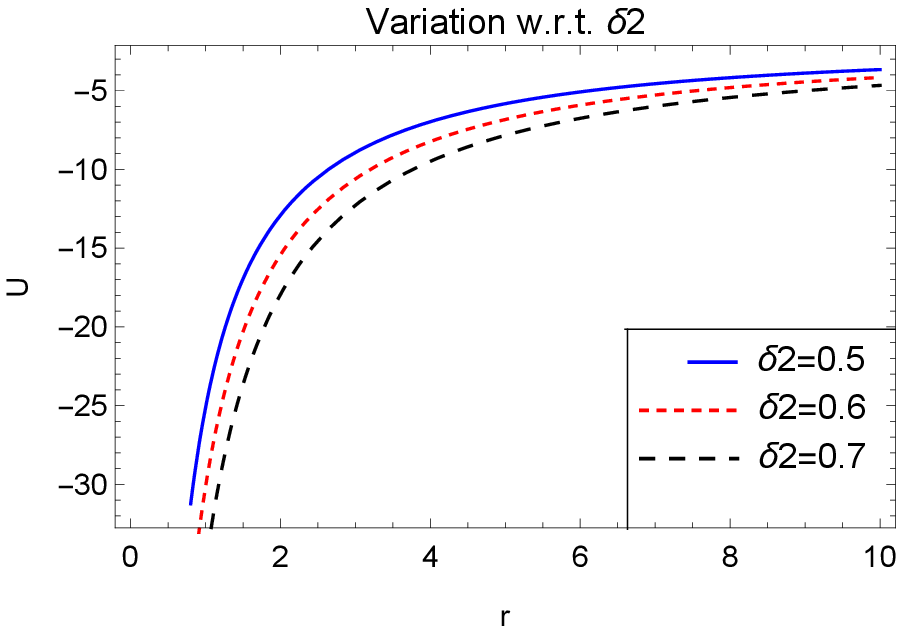}~~~~~~~\\

~~~~~~~~~~~~~~~~~~~~~~~~~~~~Fig.16~~~~~~~~~~~~~~~~~~~~~~~~~~~~~~~~~~~~~~~~~~~~~~~~~~~~~~~~~~Fig.17~~~~~~~~~\\

~~~~~~~~~~\includegraphics[height=1.7in,width=2.2in]{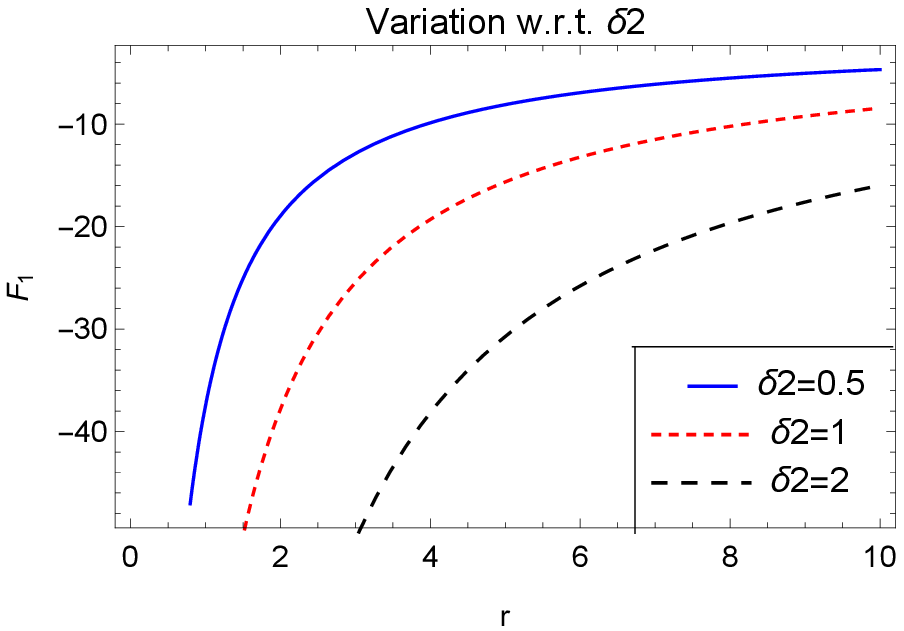}~~~~~~~~~~~~~\includegraphics[height=1.7in,width=2.2in]{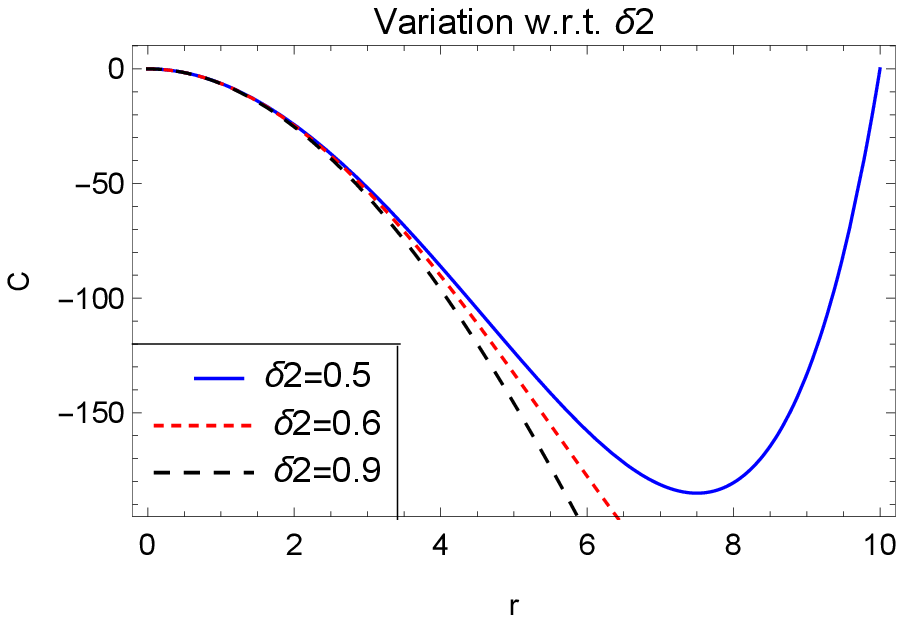}~~~~~~~\\

~~~~~~~~~~~~~~~~~~~~~~~~~~~~Fig.18~~~~~~~~~~~~~~~~~~~~~~~~~~~~~~~~~~~~~~~~~~~~~~~~~~~~~~~~~~~~~~Fig.19~~~~~~~~~\\

\vspace{1mm} \textit{\textbf{Figs.16, 17, 18 and 19} show the
variation of Thermalization temperature $T$ and Internal Energy
$U$, Helmholtz free energy $F_{1}$ and Specific Heat at constant
volume $C$ respectively with $r$ for different values of
$\delta_2$ for the Power law model (Case-1). The rainbow function
used is $\mathcal{F}(E)=\frac{1}{1-a_{1}(E_{s}/E_{P})}$. In both
the figures 16 and 17 the initial conditions are taken as
$\gamma_2=0.1$, $t=5$, $a_{1}=1$, $E_{s}=1$, $E_{P}=5$. In Fig.18
the initial conditions are $\gamma_2=0.1$, $t=5$, $a_{1}=1$,
$E_{s}=1$, $E_{P}=5$. In Fig.19 the initial conditions are
$\gamma_2=0.5$, $t=5$, $a_{1}=1$,
$E_{s}=1$, $E_{P}=5$.}\\
\end{figure}

\begin{figure}
~~~~~~~~~~~~~~~~~~~~~~~~~~\includegraphics[height=2in,width=2.5in]{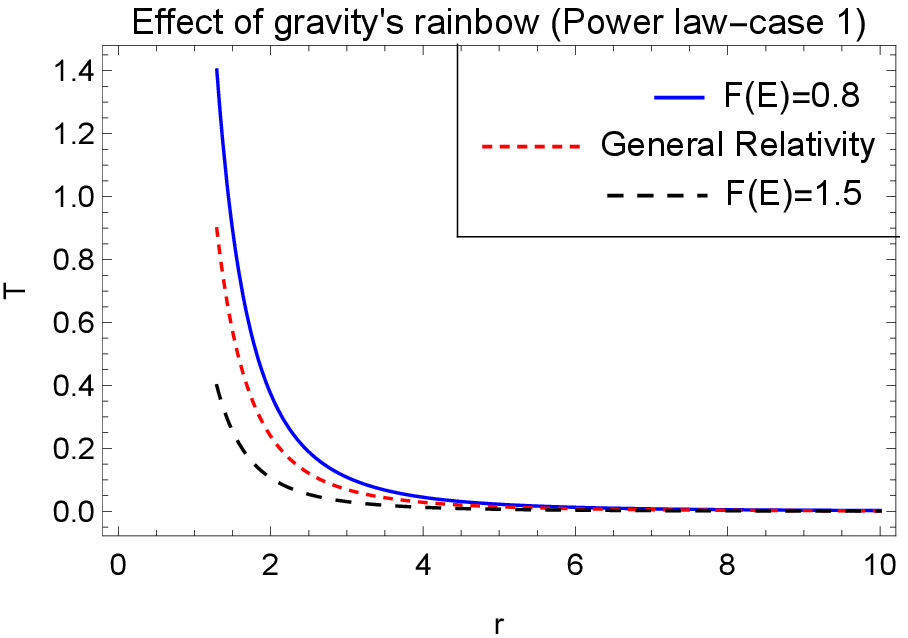}~~~~~~~~\\

~~~~~~~~~~~~~~~~~~~~~~~~~~~~~~~~~~~~~~~~~~~~~~~~~Fig.20~~~~~~~~~~~~~~~~~~\\

\vspace{1mm} \textit{\textbf{Fig.20} shows the effect of the
rainbow functions on Thermalization temperature for the Power law
model (Case-1).}
\end{figure}

In Figs.16, 17, 18 and 19, plots for the thermodynamical
parameters have been generated against $r$ for the power law model
(Case-1 $(n\geq 4)$). In Fig.16 we see that the thermalization
temperature $(T)$ decreases with the increase of $r$. Moreover
with the increase of $\delta_2$, the range of $T$ increases. From
Fig.17 we see that the internal energy $(U)$ increases with the
increase of $r$. On the contrary, with the increase of $\delta_2$,
there is decrease in the range of $U$. From Fig.18 it is evident
that Helmholtz free energy $(F_1)$ increases with $r$. Moreover
with an increase in the value of $\delta_2$, there is a
considerable decrease in the value of $F_1$. In Fig.19 the trend
of specific heat at constant volume $(C)$ is exhibited against
$r$. It is seen that with an increase in the value of $\delta_2$
there is a decrease in $C$. Finally in Fig.20, we see the effect
of the rainbow function in the thermodynamics of the system.

\subsubsection{Case-2 $(n=2)$}
The event horizon ($r_{h}$) can be obtained from the relation
$f(t,r)=0$, i.e.,
\begin{equation}\label{eventhorizon3}
-\frac{1}{\mathcal{F}^2(E)}\left[1-\frac{1}{r}\left\{A\left(f_{7}(t)-\frac{f_{8}(t)}{r}\right)+B\left(f_{9}(t)-\frac{f_{10}(t)}{r}+f_{11}(t)r^{3}+f_{12}(t)r^{4}\right)\right\}\right]=0
\end{equation}
The real positive root of the above equation gives the radius of
the event horizon. Here, the expression for thermalization
temperature becomes,
\begin{equation}\label{thermaltemp3}
T=\frac{1}{4\mathcal{F}^{2}(E)\pi
r^{3}}\left[-Arf_{7}(t)+2Af_{8}(t)-Brf_{9}(t)+2Bf_{10}(t)+2Br^{4}f_{11}(t)+3Br^{5}f_{12}(t)\right]
\end{equation}
Using the equations (\ref{thermaltemp3}), (\ref{entropy}) and
(\ref{totenergy}) we get the total energy for this model as,
\begin{equation}\label{totenergy3}
U=\frac{\pi}{2\mathcal{F}^{2}(E)}\left[-\log(r)\left\{Af_{7}(t)+Bf_{9}(t)\right\}-\frac{2}{r}\left\{Af_{8}(t)+Bf_{10}(t)\right\}+\frac{2}{3}Br^{3}f_{11}(t)+\frac{3}{4}Br^{4}f_{12}(t)\right]
\end{equation}
Using the eqns.(\ref{thermaltemp3}), (\ref{entropy}) and
(\ref{totenergy3}) we get the expression Helmholtz free energy for
this model as,
\begin{eqnarray*}
F_{1}=-\frac{\pi}{24\mathcal{F}^{2}(E)r}\left[12r\log(r)\left\{Af_{7}(t)+Bf_{9}(t)\right\}-6Arf_{7}(t)+36Af_{8}(t)-6Brf_{9}(t)\right.
\end{eqnarray*}
\begin{equation}\label{helmholtz2}
\left.+36Bf_{10}(t)+4Br^{4}f_{11}(t)+9Br^{5}f_{12}(t)\right]
\end{equation}
Finally using relations (\ref{thermaltemp3}), (\ref{entropy}),
(\ref{totenergy3}) and (\ref{specificheat}) we get the specific
heat at constant volume for this model as,
\begin{equation}\label{specificheat2}
C=\frac{\pi^{2}r^{2}\left[-Arf_{7}(t)+2Af_{8}(t)-Brf_{9}(t)+2Bf_{10}(t)+2Br^{4}f_{11}(t)+3Br^{5}f_{12}(t)\right]}{Arf_{7}(t)-3Af_{8}(t)+Brf_{9}(t)-3Bf_{10}(t)+Br^{4}f_{11}(t)+3Br^{5}f_{12}(t)}
\end{equation}
In the Figs.21, 22, 23 and 24 the thermodynamical parameters for
the power law model (Case-2 $(n=2)$) are plotted against $r$. From
Fig.21 it is evident that the thermalization temperature $(T)$
decreases with the increase in $r$. With the increase in
$\delta_3$, there is an increase in the value of $T$. From Fig.22
it is clear that there is an increase in the internal energy $(U)$
of the system with an increase in size, i.e. $r$. Moreover as
$\delta_3$ increases there is a corresponding decrease in the
value of $U$. Fig.23 shows the variation of Helmholtz free energy
$(F_1)$ against $r$. From the figure it is seen that $F_1$
increases with an increase in $r$. Further it is evident that with
an increase in $\delta_3$ there is a decrease in $F_1$. Finally in
Fig.24 we see that the specific heat at constant volume $(C)$
decreases with an increase in $r$. With an increase of $\delta_3$,
there is a corresponding increase in $C$. Finally in Fig.25 we see the effect of gravity's rainbow on the thermalization temperature.\\\\

\begin{figure}
~~~~~~~~~\includegraphics[height=1.7in,width=2.2in]{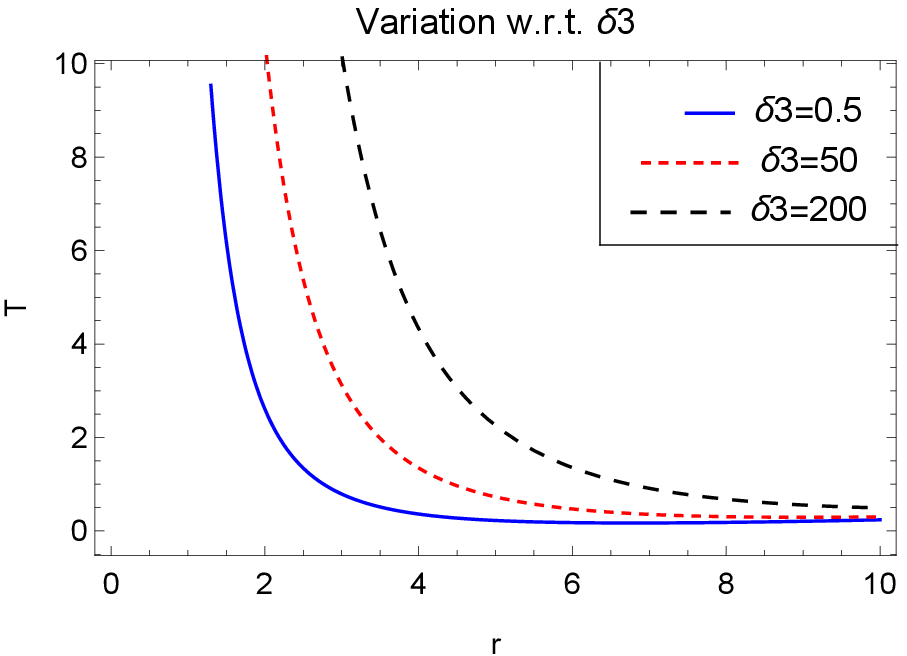}~~~~~~~~~~~~~\includegraphics[height=1.7in,width=2.2in]{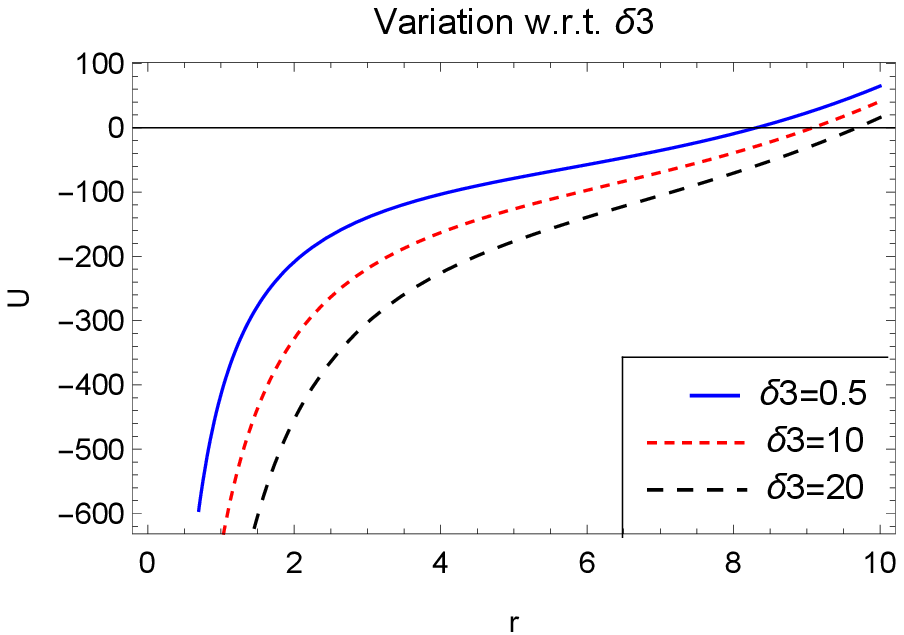}~~~~~~~\\

~~~~~~~~~~~~~~~~~~~~~~~~~~~~Fig.21~~~~~~~~~~~~~~~~~~~~~~~~~~~~~~~~~~~~~~~~~~~~~~~~~~~~~~~~~Fig.22~~~~~~~~~\\

~~~~~~~~~\includegraphics[height=1.7in,width=2.2in]{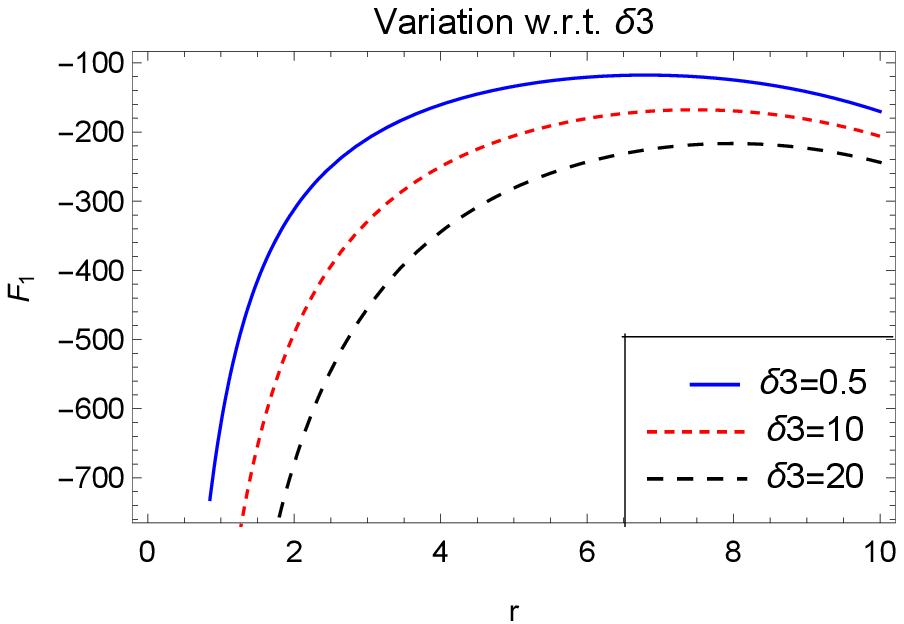}~~~~~~~~~~~~~\includegraphics[height=1.7in,width=2.2in]{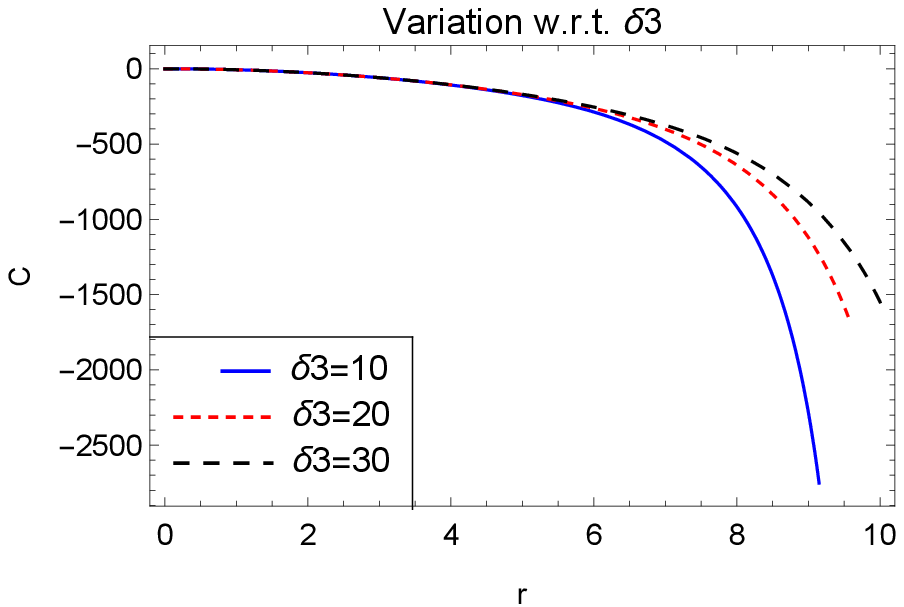}~~~~~~~\\

~~~~~~~~~~~~~~~~~~~~~~~~~~~~Fig.23~~~~~~~~~~~~~~~~~~~~~~~~~~~~~~~~~~~~~~~~~~~~~~~~~~~~~~~~~~~~~Fig.24~~~~~~~~~\\

\vspace{1mm} \textit{\textbf{Figs.21, 22, 23 and 24} show the
variation of Thermalization temperature $T$ and Internal Energy
$U$, Helmholtz free energy $F_{1}$ and Specific Heat at constant
volume $C$ respectively with $r$ for different values of
$\delta_3$ for the Power law model (Case-2). The rainbow function
is taken as $\mathcal{F}(E)=\frac{1}{1-a_{1}(E_{s}/E_{P})}$. In
both the figures 21 and 22 the initial conditions are taken as
$\gamma_3=0.1$, $\gamma_4=0.5$, $\xi_3=0.7$, $\epsilon_3=2$,
$\delta_4=10$, $A=0.5$, $B=0.8$, $t=5$, $a_{1}=1$, $E_{s}=1$,
$E_{P}=5$. In Fig.23 the initial conditions are $\gamma_3=0.1$,
$\gamma_4=0.5$, $\xi_3=0.7$, $\epsilon_3=2$, $\delta_4=10$,
$A=0.5$, $B=0.8$, $t=5$, $a_{1}=1$, $E_{s}=1$, $E_{P}=5$. In
Fig.24 the initial conditions are $\gamma_3=0.1$, $\gamma_4=0.5$,
$\xi_3=0.7$, $\epsilon_3=2$, $\delta_4=10$, $A=5$, $B=0.8$, $t=5$,
$a_{1}=1$, $E_{s}=1$, $E_{P}=5$.}

\end{figure}

\begin{figure}
~~~~~~~~~~~~~~~~~~~~~~~~~~\includegraphics[height=2in,width=2.5in]{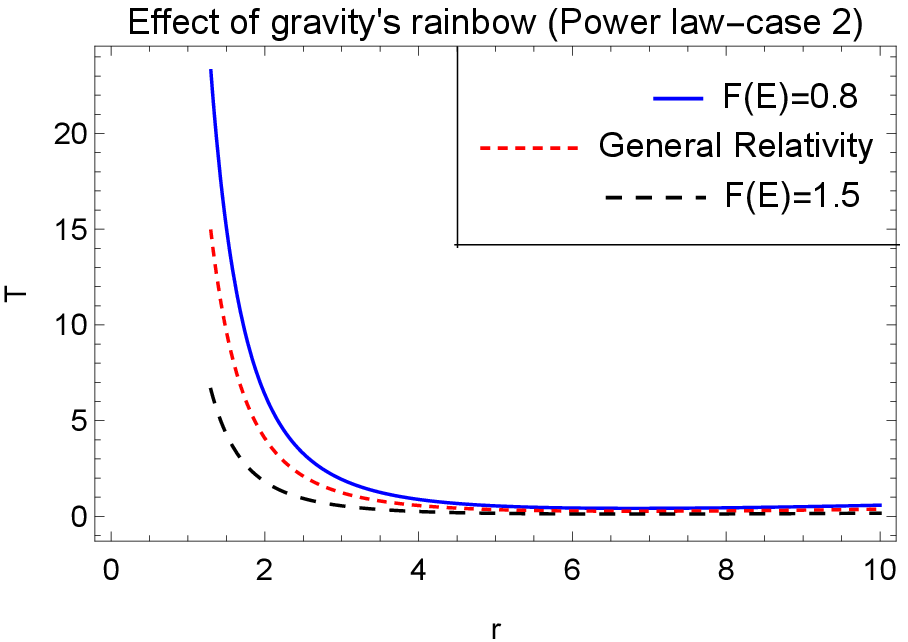}~~~~~~~~\\

~~~~~~~~~~~~~~~~~~~~~~~~~~~~~~~~~~~~~~~~~~~~~~~~~~~~Fig.25~~~~~~~~~~~~~~~~~~\\

\vspace{1mm} \textit{\textbf{Fig.25} shows the effect of the
rainbow functions on Thermalization temperature for the Power law
model (Case-2).}
\end{figure}

\section{Conclusion and Discussion}
In this work we have investigated rainbow modified $f(R)$ gravity
via a gravitational collapse and thermodynamical study. We have
considered a non-static (time dependent) geometry, namely the
Vaidya spacetime for our study. The collapsing procedure being a
time dependent phenomenon justifies our choice. Rainbow
modifications to Vaidya geometry was considered, thus making the
spacetime energy dependent in nature. As discussed earlier the
rainbow modifications to a metric is significant only in the
quantum regime, i.e. when the energy of the particle probed is
comparable to the Planck energy $(E_P\approx10^{19} GeV)$ and the
length scale tends to the Planck length $(L_P\approx10^{-33} cm)$.
Under such a scenario the quantum fluctuations will be at play,
and will have far reaching effect on the dynamics of the system.
This was the motivation of our model and we wanted to probe such
effects via a theoretical framework.

The field equations for such a system was derived for the general
$f(R)$ expression. In order to get a solution of the system, we
needed to consider various models of $f(R)$ gravity. In this work
we considered the famous inflationary Starobinsky's model and the
Power law model of $f(R)$ gravity with suitable motivations.
Solutions for each of these models were obtained separately which
gave the mass of the collapsing system as a function of both the
radial and temporal coordinates. Using them in the metric we
obtained the geometries for each of the models. Here it should be
mentioned that in the process of solving the system we have used
the $(11)$ component of the field equations given by
eqn.(\ref{11}). This has been done purely due to the simplicity of
the equation compared to the others with no other bias. This
choice helped us to reduce our computational efforts considerably,
but in doing so we have missed out on a very important parameter,
namely the density parameter $\rho$ of the matter content of the
system. This is solely because $\rho$ did not feature in the
$(11)$ component of the field equations and hence remained absent
in the subsequent analysis. So in our analysis dependence of the
collapsing procedure on the matter content of the system could not
be studied, which in turn restricted us from mapping our results
to various cosmological eras.

A geodesic study was performed in order to study the collapsing
procedure of the star in both the models. The study was aimed at
enquiring the nature of singularity resulting from the collapse. A
singularity censored by an event horizon will be a BH, but any
singularity devoid of such horizon will be a naked one and easily
accessible from the outside. So the loss of information and other
problems posed by a BH will cease to exist for such type of
singularities. This motivated us to initiate such a study. However
the complexity of the systems provoked us to use numerical
analysis in our study. In this context we mention here that our
choice for the functions $f_{i}(t)$, $i=1,2,3...12$ seems to be
self similar in nature. This has been done in accordance with the
definition of $X_{0}$ in eqn.(\ref{X0}), such that the gradient
$t/r$ can be formed without much complications. However, naturally
the question arises that what would happen if we consider non
self-similar terms? From the mathematical framework we see that
under the limit $r\rightarrow 0, t\rightarrow 0$ non self-similar
terms will result in the deletion of crucial terms or in the
creation of mathematically undefined terms, both of which are
undesirable. Deletion of important terms will result in loss of
information about the system while undefined terms will make
further mathematical computations next to impossible. From the
study it is seen that NS is quite a possibility along with BH for
the collapsing models. Fine tuning the initial conditions we may
get NS or BH for different ranges of the parameters involved. It
should be mentioned that this is a significant counter-example of
the cosmic censorship hypothesis. Moreover it is seen that the
naked singularity formed can be globally naked for certain
scenarios.

An important result that we have derived from this study is that
under the effect of gravity's rainbow there is greater tendency of
formation of NS. The deviations produced by the presence of
rainbow functions favours the formation of NS. This is a very
interesting finding given the renewed interest of scientists in
NS. It is admitted that the result obtained depends on the initial
data, which is expected from previous works like \cite{cch3, ini}.
Over the years it has been seen that the role of initial data is
central to any study of gravitational collapse. We know that we
have not been able to detect a NS till date. In the light of such
a comment it seems that our work loses some significance. But on
the contrary we would like to emphasize on the fact that our work
should generate widespread motivations for the search of NS. This
is more so because for quite sometime it is strongly believed that
with the failure of cosmic censorship we would come face to face
with the laws of quantum gravity, whenever gravitational collapse
of a distant star results in a NS. Thus a complete Theory Of
Everything (TOE) needs a proper knowledge of NS, which can act as
a laboratory for quantum gravity.

We have also investigated the conditions for a strong singularity
for different models. The conditions have been derived and
supplemented with suitable examples where required. Finally we
have concluded our analysis with a thermodynamical study of the
system. Expressions for various thermodynamical parameters like
the thermalization temperature, Internal energy, etc. have been
deduced for the models and their nature have been analyzed via
various plots. In a classical set-up it is expected that the
thermalization temperature $(T)$ should decrease with distance
from the core of the star or with an increase in the size of the
star. From the definition of the Internal energy $(U)$ it is clear
that its value should increase with the size of the star for a
given $T$. Similarly the Helmholtz free energy $F_1$ should also
increase with an increase in $r$, given its definition and
relative trends of $T$ and $U$. Our results show deviations from
the above expected results. Such deviations can be attributed to
the quantum fluctuations of our model. Our study reveals
information that characterizes a quantum evolution of the
universe. So we are hopeful on the fact that these results will be
useful in any future attempts towards the formulation of a
successful theory of quantum gravity.

%%%%%%%%%%%%%%%%%%%%%%%%%%
\section*{Acknowledgments}
%%%%%%%%%%%%%%%%%%%%%%%%%%

The author acknowledges the Inter University Centre for Astronomy
and Astrophysics (IUCAA), Pune, India for granting visiting
associateship.

\section{Appendix}
Here we report the other components of the field equations for
this model which have not been used in our analysis.\\

\textbf{\textit{1. The (00)-component of field equations is given
by,}}
\begin{eqnarray*}
f''(R)\mathcal{G}^{2}(E)\left\{-2\mathcal{F}^{2}(E)r^{4}\left(2\ddot{m}'+\ddot{m}''r\right)+\mathcal{G}^{2}(E)
\left(m-r\right)\left(m-m'r\right)\left(-4m'+r\left(m''+m'''r\right)\right)\right.
\end{eqnarray*}
\begin{eqnarray*}
\left.-\mathcal{F}(E)\mathcal{G}(E)r^{2}\left(m\dot{r}\left(m''+m'''r\right)
-m\left(2\dot{m}'+\dot{m}''r\right)+m'\left(-4\dot{m}+r\left(2\dot{m}'+\dot{m}''r\right)\right)\right)\right\}
\end{eqnarray*}
\begin{eqnarray*}
-r^{2}\left\{-2\mathcal{F}(E)\mathcal{G}(E)f'(R)\dot{m}r^{2}+2\mathcal{F}^{2}(E)\mathcal{G}^{4}(E)f'''(R)\left(2\dot{m}'+\dot{m}''r\right)^{2}
+r\left(r-m\right)\left(f'(R)\left(-2\mathcal{G}^{2}(E)m'+r^{2}R\right)\right.\right.
\end{eqnarray*}
\begin{equation}\label{00}
\left.\left.+r^{2}\left(2~\Box
f'(R)-f(R)+2\kappa\rho\right)\right)+2r^{4}\kappa\sigma\right\}=0
\end{equation}

\textbf{\textit{where $\Box f'(R)$ is given by,}}
\begin{eqnarray*}
\Box
f'(R)=\frac{1}{r^{7}}\left[f'''(R)\mathcal{G}^{5}(E)\left\{4m'-r\left(m''+rm'''\right)\right\}\left\{-2\mathcal{F}(E)r^{2}
\left(2\dot{m}'+r\dot{m}''\right)+\mathcal{G}(E)\left(m-r\right)
\right.\right.
\end{eqnarray*}
\begin{eqnarray*}
\left.\left.\left(-4m'+r\left(m''+rm'''\right)\right)\right\}+f''(R)\mathcal{G}^{3}(E)r^{2}\left\{2r^{2}\mathcal{F}(E)\left(-2\dot{m}'
+r\left(2\dot{m}''+\dot{m}'''r\right)\right)\right.\right.
\end{eqnarray*}
\begin{eqnarray*}
\left.\left.+\mathcal{G}(E)\left(-m\left(8m'+r\left(-5m''+r^{2}m''''\right)\right)+r\left(4m'^{2}-m'\left(-4+r\left(m''+rm'''\right)\right)\right.\right.\right.\right.
\end{eqnarray*}
\begin{equation}\label{dalembert}
\left.\left.\left.\left.+r\left(-4m''+r\left(m'''+rm''''\right)+rm'''\right)\right)\right)\right\}\right]
\end{equation}
\\

\textbf{\textit{2. The (22)-component of field equations is given
by,}}
\begin{eqnarray*}
-2f'''(R)\mathcal{G}^{5}(E)\left\{4m'-r\left(m''+m'''r\right)\right\}\left[-2\mathcal{F}(E)r^{2}\left(2\dot{m}'+\dot{m}''r\right)+\mathcal{G}(E)
\left(m-r\right)\left\{-4m'+r\left(m''+m'''r\right)\right\}\right]
\end{eqnarray*}
\begin{eqnarray*}
+r^{2}\left[2f''(R)\mathcal{G}^{3}(E)\left\{\mathcal{F}(E)r^{2}\left(6\dot{m}'-r\left(3\dot{m}''+2\dot{m}'''r\right)\right)
+\mathcal{G}\left(r\left(-4m'^{2}+m'\left(-8+r\left(m''+m'''r\right)\right)-r\left(-5m''\right.\right.\right.\right.\right.
\end{eqnarray*}
\begin{equation}\label{22}
\left.\left.\left.\left.\left.+m^{iv}r^{2}+rm'''\right)\right)
+m\left(12m'+r\left(-6m''-m'''r+m^{iv}r^{2}+rm'''\right)\right)\right)\right\}+r^{5}\left(f(R)-f'(R)R+2\kappa
\omega\rho\right)\right]=0
\end{equation}
\\

\textbf{\textit{3. The (33)-component of field equations is given
by,}}
\begin{eqnarray*}
-2f'''(R)\mathcal{G}^{5}(E)\left\{4m'-r\left(m''+m'''r\right)\right\}\left[-2\mathcal{F}(E)r^{2}\left(2\dot{m}'+\dot{m}''r\right)+\mathcal{G}
\left(m-r\right)\left\{-4m'+r\left(m''+m'''r\right)\right\}\right]
\end{eqnarray*}
\begin{eqnarray*}
+r^{2}\left[2f''(R)\mathcal{G}^{3}(E)\left\{12\mathcal{G}(E)m
m'-2\mathcal{G}(E)\left(3mm''+2m'\left(2+m'\right)\right)r+\left(\mathcal{G}m''\left(m'+5\right)+6\mathcal{F}(E)\dot{m}'\right)r^{2}+\left(\mathcal{G}(E)\right.\right.\right.
\end{eqnarray*}
\begin{equation}\label{33}
\left.\left.\left.\left(mm^{iv}+m'''\left(m'-1\right)\right)-3\mathcal{F}(E)\dot{m}''\right)r^{3}-\left(\mathcal{G}(E)m^{iv}
+2\mathcal{F}(E)\dot{m}'''\right)r^{4}\right\}+r^{5}\left(f(R)-f'(R)R+2\kappa\omega\rho\right)\right]=0
\end{equation}
\\

\textbf{\textit{4. The $(01)$ or $(10)$-component of field
equations is given by,}}
\begin{eqnarray*}
2f'''(R)\mathcal{G}^{5}(E)\left\{4m'-r\left(m''+m'''r\right)\right\}\left[\mathcal{F}(E)r^{2}\left(2\dot{m}'+\dot{m}''r\right)-\mathcal{G}(E)\left(m-r\right)
\left\{-4m'+r\left(m''+m'''r\right)\right\}\right]
\end{eqnarray*}
\begin{eqnarray*}
+r^{2}\left[f''(R)\mathcal{G}^{3}(E)\left\{-2\mathcal{F}(E)r^{3}\left(3\dot{m}''+\dot{m}'''r\right)+\mathcal{G}(E)\left(r\left(-4m'^{2}
+m'\left(-8+r\left(m''+m'''r\right)\right)-2r\left(-4m''\right.\right.\right.\right.\right.
\end{eqnarray*}
\begin{eqnarray*}
\left.\left.\left.\left.\left.+r\left(m'''+m^{iv}r\right)+rm'''\right)\right)+m\left(12m'+r\left(-9m''+r\left(m'''+2m^{iv}r\right)+2rm'''\right)\right)\right)\right\}
+r^{5}\left(f(R)\right.\right.
\end{eqnarray*}
\begin{equation}\label{01}
\left.\left.-f'(R)R-2\kappa\rho\right)\right]=0
\end{equation}

%%%%%%%%%%%%%%%%%%%%%%%%%%%%%%%%%%%%%%%%%%%%%%%%%%%%%%%%%%%%%%%%%%%%%
%%%%%%%%%%%%%%%%%%%%%%%%%%%%%%%%%%%%%%%%%%%%%%%%%%%%%%%%%%%%%%%%%5

\end{document}